\documentclass[conference]{IEEEtran}
\IEEEoverridecommandlockouts
\usepackage{cite}
\usepackage{array}
\usepackage{tabularx}
\usepackage{amsmath,amssymb,amsfonts}
\usepackage{algorithmic}
\usepackage{graphicx}
\usepackage{textcomp}
\usepackage{xcolor}
\usepackage{booktabs}
\usepackage{multirow}
\usepackage{booktabs}
\usepackage[table]{xcolor}
\usepackage{amsfonts}
\usepackage{hyperref}
\usepackage{cleveref}
\usepackage{tikz}

\usepackage{amsmath,graphicx,hyperref}
\usepackage{cite}
\usepackage{multirow}
\usepackage{url} 
\usepackage{amsmath,amssymb,amsfonts}
\usepackage{algorithmic}
\usepackage{adjustbox}
\usepackage{textcomp}
\usepackage{booktabs}
\usepackage[table]{xcolor}

\definecolor{lightgray}{RGB}{240,240,240}
\definecolor{lightblue}{RGB}{240,248,255}
\definecolor{lightyellow}{RGB}{255,250,205}

\definecolor{lightgreen}{HTML}{EAF2E3}
\definecolor{lightorange}{HTML}{FFF3E6}

\definecolor{lightgray}{RGB}{240,240,240}
\definecolor{lightblue}{RGB}{240,248,255}
\definecolor{lightyellow}{RGB}{255,250,205}

\definecolor{lightgreen}{HTML}{EAF2E3}
\definecolor{lightorange}{HTML}{FFF3E6}
\def\BibTeX{{\rm B\kern-.05em{\sc i\kern-.025em b}\kern-.08em
    T\kern-.1667em\lower.7ex\hbox{E}\kern-.125emX}}
\begin{document}

\title{
The Sound of Absence: Audio-Language Embedding Models Struggle with Negation
}



\author{
Chun-Yi Kuan$^{\heartsuit}$, 
Hung-yi Lee$^{\heartsuit}$$^{\clubsuit}$ \\
$^{\heartsuit}$Graduate Institute of Communication Engineering, National Taiwan University, Taiwan  \\
$^{\clubsuit}$Artificial Intelligence Center of Research Excellence (AI-CoRE), National Taiwan University, Taiwan \\
chunyi.kuan.tw@gmail.com
}





\maketitle

\begin{abstract}
Audio-language embedding models such as CLAP are widely evaluated on matching present sound events, but rarely on negation. 
We show this affirmation-only evaluation hides a key limitation: these models fail to encode negated sound concepts, mapping affirmative and negated captions to nearly identical representations. 
To expose this blind spot, we introduce NegEval-Audio, a framework that converts existing datasets into two negation-aware tasks, Retrieval-Neg and Multiple-Choice Negation (MCQ-Neg), to probe whether models distinguish present from absent events. 
On AudioCaps and Clotho, performance degrades sharply under negation, with negation-type MCQ accuracy falling far below chance, and the failure persists even for a recent multimodal LLM-based embedding model. 
While a training-free steering method improves MCQ-Neg, it yields marginal gains for Retrieval-Neg. 
This indicates that affirmation bias is a fundamental flaw in the representation geometry, necessitating explicit negation-aware training objectives.
\end{abstract}


\begin{figure*}[ht]
    \centering
    \includegraphics[width=0.95\textwidth]{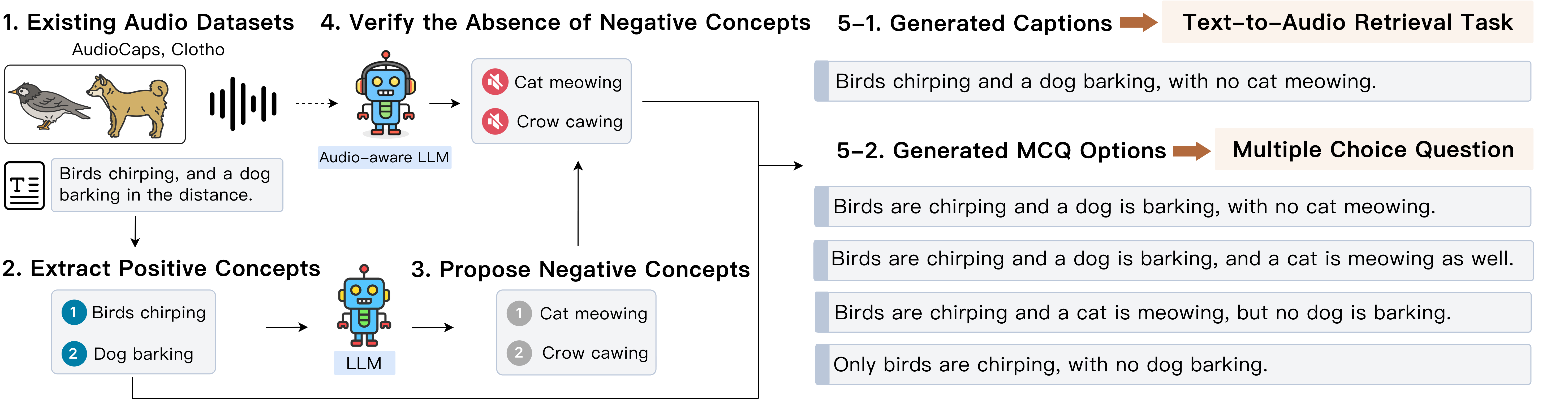} 
    \caption{
    Overview of NegEval-Audio. 
    Existing audio-caption datasets are converted into negation-aware retrieval and multiple-choice tasks by extracting present sound concepts and verifying plausible absent concepts.
    }
    \label{fig:overview}
\end{figure*}
\section{Introduction}
Recent progress in audio-language embedding models, which map audio clips and text into a shared space for retrieval and matching, has enabled strong performance on benchmarks such as AudioCaps~\cite{kim2019audiocaps} and Clotho~\cite{drossos2020clotho}. 
This family now spans contrastive bi-encoders such as CLAP~\cite{wu2023large, elizalde2023clap, ye2023clapspeech, yuan2024t, jing2024paraclap, li2024advancing, ghoshcompa, takano2025human, tseng2025revisiting, dinkel2026glap} and M2D-CLAP~\cite{niizumi2024m2d-clap,
niizumi2025m2d-clap}, as well as recent embedding models built on multimodal LLMs such as WAVE~\cite{tang2025wave}. 
As a result, modern models have become increasingly effective at detecting and matching acoustic events that are present. 
However, these evaluations share an implicit assumption: audio semantics is affirmation-only. 
Models are rewarded for recognizing sounds that occur, but are rarely examined on whether they understand which plausible
sounds do not occur. 
This blind spot matters: in natural language, negation is fundamental, and many practical audio–language applications depend on it. 
For example, a user may search for queries such as ``rain without thunder'' or ``speech without background music'', which are common, precise, and meaningful in practice.
Despite this importance, negation remains relatively underexplored in audio–language embedding models.
Current embedding models are naturally trained to associate audio with sounds that are present, but not to account for sounds that a description explicitly negates.

To address this gap, we introduce \emph{NegEval-Audio}, an evaluation framework designed to probe negation understanding in audio–language embedding models.
Figure~\ref{fig:overview} provides an overview. 
Instead of constructing a new dataset from scratch, our framework transforms existing audio retrieval datasets into negation-aware evaluation sets. 
We first extract positive (existent) sound
concepts from the caption and then propose corresponding negative (non-existent) concepts. 
These negative concepts are verified using audio-aware large language
models (ALLMs)~\cite{chu2024qwen2, gong2023joint, comanici2025gemini,
achiam2023gpt, xu2025qwen2, abouelenin2025phi, tang2023salmonn, chu2023qwen, gong2023listen, ghoshaudio, goel2025audio, yang2024building, chang2024speechprompt, fathullah2023towards, wang2023blsp, wang2024blsp, ji2024wavchat, held2025distilling, peng2025survey, cui2025recent, mousavidiscrete, wu2024towards, guo2025recent, kuan2024speech, kuan2025teaching, kuan2025alignment, arora2025landscape, chang2026tico}, ensuring that they truly do not occur in the audio. This pipeline allows standard retrieval datasets to be directly converted into negation-aware formats without additional data collection. 
Based on the verified concept pairs, we construct two evaluation tasks. The first is text-to-audio retrieval with negation (Retrieval-Neg), which tests whether models can handle realistic queries that combine affirmative and negative statements, e.g., ``Birds are chirping and rain is falling, with no human voices in the background.'' 
This requires the model not only to retrieve audio containing specific events, but also to ensure that other events are absent, unlike traditional retrieval, which considers only presence. 
The second is multiple-choice questions with negation (MCQ-Neg), whose carefully designed options reveal specific failure types: given several closely related descriptions, the model must select the one that correctly reflects the audio, while distractors differ only by affirming or denying particular events. 
For instance, ``A baby is crying with no adult voices present'' and ``Adult voices are present with no baby crying'' are surface-similar yet imply opposite judgments, requiring attention to subtle but meaningful differences.

Through this protocol, we find that audio--language embedding models struggle with negation. In the embedding space, affirmative and negated captions often occupy highly overlapping regions; for example, ``a sound of cat meowing'' and ``no cat meowing sound'' receive similar representations.
The model thus fails to encode the distinction between the presence and the explicit absence of a sound event, revealing a persistent affirmation bias that holds even for the recent multimodal LLM-based WAVE~\cite{tang2025wave}. 
Recognizing this limitation, we further explore mitigation strategies. 
Rather than the obvious data-centric route of augmenting training captions with synthetic negation, we investigate a lightweight, training-free steering technique to adjust the embedding space directly. 
While this intervention improves performance on MCQ-Neg, it yields only marginal gains for Retrieval-Neg. 
This contrast reveals that the models' inability to handle negation is not a mere surface-level artifact, but a fundamental flaw in their representation geometry, underscoring the need for explicit negation-aware training objectives.
Our findings indicate that robust audio understanding has to go beyond recognizing present events and incorporate explicit reasoning about negation and sound absence. 


Our contributions are summarized as follows:
(1) We identify negation as a critical and previously overlooked failure mode in audio–language embedding models.
(2) We introduce NegEval-Audio, a systematic framework that transforms existing audio-caption datasets into negation-aware retrieval and multiple-choice benchmarks. 
Across CLAP, M2D-CLAP, and a recent MLLM-based embedding model, we show that current models fail to reliably distinguish present from absent sound events.
(3) We provide a controlled text-side diagnosis showing that negated descriptions remain close to both their affirmative counterparts and bag-of-concepts representations, and explore training-free embedding steering as a lightweight mitigation. 
Steering improves local option ranking in MCQ-Neg but yields only marginal retrieval gains, indicating a deeper limitation in the joint embedding geometry.


\section{Related Work}

\subsection{Audio--language embedding models.}
Contrastive audio--language models~\cite{wu2023large, elizalde2023clap, yuan2024t, li2024advancing, ghoshcompa, niizumi2024m2d-clap, niizumi2025m2d-clap, tang2025wave} learn a shared space between audio clips and natural language descriptions, enabling text-to-audio retrieval, audio--text matching, and zero-shot audio classification. Representative models such as CLAP~\cite{wu2023large, elizalde2023clap} and its variants~\cite{ye2023clapspeech, yuan2024t, jing2024paraclap, li2024advancing, ghoshcompa, niizumi2024m2d-clap, takano2025human, tseng2025revisiting, niizumi2025m2d-clap, dinkel2026glap} have shown strong performance on standard benchmarks such as AudioCaps~\cite{kim2019audiocaps} and Clotho~\cite{drossos2020clotho}. 
Recent benchmarks further examine compositional audio understanding, including multiple sound events, attributes, and temporal relations. 
However, these evaluations are still largely built around affirmative captions that describe sounds present in the audio. 
Whether audio--language embedding models can also handle descriptions that specify what should be absent remains underexplored.

\subsection{Compositionality and negation in multimodal embeddings.}
Prior work~\cite{yuksekgonuland, thrush2022winoground, zhao2022vl, ma2023crepe, hsieh2023sugarcrepe} in vision--language modeling has shown that contrastive embeddings often rely on salient content words while underusing relational structure, word order, and logical operators. 
This behavior is commonly described as bag-of-words-like matching~\cite{lewis2024does}: models recognize individual concepts but may fail to represent how these concepts are composed. 
Negation is a representative failure case. 
Vision--language models~\cite{alhamoud2025vision, zhang2025negvqa, seeing_negation_2026} can assign similar representations to affirmative and negated descriptions, such as ``a dog'' and ``no dog,'' revealing an affirmation bias in the joint embedding space. 
These findings suggest that strong retrieval performance does not necessarily imply robust logical understanding. 
Motivated by this observation, we ask whether a similar failure mode appears in audio--language embeddings, where queries such as ``speech without background music'' and ``rain without thunder'' are natural and practically meaningful.



\section{NegEval-Audio Benchmark Construction}

\subsection{Overview of NegEval-Audio Construction Pipeline}

Existing audio-language datasets are typically built under an affirmation-only assumption, where captions describe sound events that are present in the audio.
Our key idea is to transform these standard datasets into negation-aware evaluation sets without collecting new audio.
Instead of creating new annotations from scratch, we derive negation semantics from the audio-caption pair itself, using captions to identify present sound events and audio-side verification to confirm plausible absent ones.
Formally, let an audio sample be denoted as $a$, paired with its caption $c$.
From $c$, we extract a set of \textbf{positive sound concepts} $\mathcal{P}(a)=\{p_1, p_2, \dots, p_m\}$,
where each $p_i$ corresponds to a sound event explicitly described as occurring in the audio.
We then construct a complementary set of \textbf{negative sound concepts} $\mathcal{N}(a)=\{n_1, n_2, \dots, n_k\}$,
where each $n_j$ represents a sound event that is plausible in the same acoustic scene but does \emph{not} occur in the audio.
These negative concepts are first proposed by an LLM based on contextual plausibility inferred from the caption, and are then verified using audio-aware LLMs to ensure their absence.
Through this process, each audio sample is associated with a pair of concept sets $(\mathcal{P}(a), \mathcal{N}(a))$, representing what the audio contains and what it explicitly does not contain.
This concept-level representation enables us to systematically construct two negation-aware evaluation tasks:
\textbf{(1) Retrieval-Neg}, where text queries include both positive and negative statements, and
\textbf{(2) MCQ-Neg}, where closely related descriptions differ only by affirming or negating specific sound events.
Importantly, NegEval-Audio is dataset-agnostic and can convert existing affirmation-only audio-language retrieval benchmarks into negation-aware evaluation sets.
In the following subsections, we describe how positive concepts are extracted from captions, how negative concepts are proposed and verified, and how these concept sets are used to construct the two tasks.

\subsection{Positive Concept Extraction from Captions}

Given an audio sample $a$ and its paired caption $c$, the first step of our pipeline is to extract the set of sound events that are explicitly described as occurring in the audio. 
We employ an LLM to perform semantic extraction of sound events from captions. 
The LLM is prompted to identify and normalize the sound events described in $c$ into concise concept phrases that represent distinct acoustic events. For example, the caption ``A dog barking while rain is falling in the background'' is mapped to
$\mathcal{P}(a) = \{\text{dog barking}, \text{rain falling}\}.$
This step preserves the semantic intent of the caption while converting it into a structured concept-level representation.
The extracted positive concepts serve as the semantic anchor of each audio sample and form the basis for constructing negation-aware evaluation tasks in subsequent stages.


\subsection{Negative Concept Proposal and Verification}

After obtaining the positive concept set $\mathcal{P}(a)$ from the caption, the next step is to construct a complementary set of \textbf{negative sound concepts}. 
We first denote the set of candidate negatives proposed by the LLM as
$\tilde{\mathcal{N}}(a) = \{\tilde{n}_1, \tilde{n}_2, \dots, \tilde{n}_k\}$,
where each $\tilde{n}_j$ represents a sound event that is plausible in the same acoustic scene but not mentioned in the caption.
A naive approach would be to randomly select sound events as negatives. 
However, this would lead to unrealistic negation cases that do not reflect meaningful semantic reasoning. 
Instead, we require candidate negatives to satisfy a \textit{contextual plausibility} constraint. 
Given the positive concept set $\mathcal{P}(a)$, we prompt the LLM to suggest sound events that could reasonably co-occur with these concepts in the same acoustic scene.
For example, given the caption ``A dog barking while rain is falling'', the LLM might propose candidates like \textit{human speech}, \textit{car passing}, or \textit{thunder}.

However, plausibility inferred from text alone is insufficient.
We therefore introduce an audio-side verification step using an audio-aware LLM.
For each candidate concept $\tilde{n}_j \in \tilde{\mathcal{N}}(a)$, we query the audio-aware LLM with a sound-existence question of the form: \textit{``Does the audio contain the sound of $\tilde{n}_j$?''}
Its response is then used to determine whether $\tilde{n}_j$ is present in the audio.
To assess the reliability of the audio-aware LLM verification step, we manually validated a subset of the constructed negative concepts.
We randomly sampled 200 audio clips from AudioCaps and Clotho, covering 460 automatically verified negative concepts. 
A human annotator judged whether each negative concept verified by the audio-aware LLM was indeed absent from the corresponding audio.
The annotator agreed with the audio-aware LLM verification in 445 out of 460 cases, yielding a 96.7\% agreement rate.
Only the candidates verified as absent are retained to form the final negative concept set:
$\mathcal{N}_{v}(a) = \{ n_j \mid \tilde{n}_j \in \tilde{\mathcal{N}}(a), \ \textit{Audio-aware LLM indicates absence in } a \}$.

Note that this verification step does not presuppose the capability under evaluation. 
The audio-aware LLM is only asked affirmative sound-existence questions (i.e., whether a sound is present), a task on which such models perform reliably; it is never required to interpret negated language. 
The negation semantics are introduced afterward, at the text-construction stage. 
Our benchmark then evaluates embedding models, a different model family, on these negated descriptions. The 96.7\% human agreement further confirms the reliability of this step.

Through this two-stage process, negative concepts are \textbf{text-proposed} and \textbf{audio-verified}. 
This ensures that $\mathcal{N}_{v}(a)$ consists of realistic, semantically meaningful, and truly absent sound events, rather than arbitrary negatives.
The resulting pair $(\mathcal{P}(a), \mathcal{N}_{v}(a))$ provides a structured semantic representation of what the audio contains and what it explicitly does not contain.

\subsection{Construction of Retrieval-Neg Task}

Using the verified concept sets $(\mathcal{P}(a), \mathcal{N}_{v}(a))$, we transform the original standard retrieval setting into a negation-aware task.
For each audio sample $a$, we construct a negation-aware query $q_{\text{neg}}$ that contains both positive and negative statements. 
Conceptually, the query is designed to satisfy two semantic constraints: the audio must contain $\mathcal{P}(a)$ and must not contain $\mathcal{N}_{v}(a)$.
This changes the nature of the retrieval problem from affirmation-only matching to negation-constrained matching, while keeping the retrieval protocol unchanged.

In practice, we construct a negation-aware query $q_{\text{neg}}$ using a simple rule-based template to ensure consistency.
For each audio sample, we start from the original caption $c$ and append a negation statement using one sampled concept from $\mathcal{N}_{v}(a)$. 
The template takes the form: original caption $c$ + \textit{``There is no sound of \{absent sound\}''}, where \{absent sound\} is sampled from the verified negative concept set $\mathcal{N}_{v}(a)$. 
For example, suppose the original caption is \textit{``A dog barking while rain is falling''} and a verified negative concept is \textit{human speech}. 
The rule-based construction yields \textit{``A dog barking while rain is falling. There is no sound of human speech''}.
Since this template-based sentence may sound slightly unnatural, we further apply the LLM to paraphrase the query into a more fluent form without altering its semantics. 
The above example becomes \textit{``A dog barking while rain is falling, with no human speech''}.
This two-step process ensures that negation queries are generated in a controlled manner, while maintaining natural linguistic quality for retrieval evaluation.
Importantly, this task remains within the standard retrieval framework. 
The only change lies in the semantic structure of the query. 
This allows us to directly compare model performance between standard retrieval and \textit{Retrieval-Neg}.

\subsection{Construction of MCQ-Neg Task}

While Retrieval-Neg evaluates coarse-grained negation in a retrieval setting, MCQ-Neg provides a sentence-level diagnostic of how models interpret negation. 
For each audio $a$, we construct a four-option multiple-choice question derived from the structured pair $(\mathcal{P}(a), \mathcal{N}_v(a))$, where candidate sentences differ only in the affirmation or negation of specific sound events.
Following prior work~\cite{laka1990negation, alhamoud2025vision}, we categorize MCQs into three types based on the structure of the correct answer: \textit{Affirmation}, \textit{Negation}, and \textit{Hybrid}. 
Each type probes distinct aspects of negation and exposes different failure modes.
Options are generated programmatically using prompt-encoded logical templates with an LLM. 
Given $\mathcal{P}(a)$ and $\mathcal{N}_v(a)$, the model produces one correct option and three distractors according to predefined logical rules (e.g., false affirmation, reversed logic, false exclusion). 
For each MCQ type, we specify dedicated prompts that enforce exact logical constraints.
\noindent\textbf{Affirmation MCQ.}  
The correct option describes only $\mathcal{P}(a)$. 
Distractors introduce incorrect events from $\mathcal{N}_v(a)$ or omit true ones.
\noindent\textbf{Negation MCQ.}  
The correct option negates a concept from $\mathcal{N}_v(a)$ while remaining consistent with $\mathcal{P}(a)$. 
Distractors invert this logic.
\noindent\textbf{Hybrid MCQ.}  
The correct option simultaneously affirms $\mathcal{P}(a)$ and negates $\mathcal{N}_v(a)$, while distractors minimally alter this structure to create semantically opposite statements. 
For example, if $\mathcal{P}(a) = \{\text{dog barking}, \text{rain}\}$ and $\mathcal{N}_v(a) = \{\text{human speech}\}$, a generated Hybrid MCQ may include:
(A) A dog is barking while rain is falling, with no human speech present \textit{(correct)};  
(B) A dog is barking while rain is falling, and human speech can be heard \textit{(distractor)};  
with additional options constructed by systematically reversing affirmation and negation.
All generated options are lightly paraphrased to improve naturalness without changing the underlying sound concepts. 
A subset of the generated MCQs was manually verified to ensure logical correctness and linguistic validity. 

\section{Training-Free Negation Steering}

Beyond diagnosis, we further explore whether negation bias can be mitigated.
Inspired by prior work~\cite{seeing_negation_2026, pai2025billy, xie2025emosteer, chen2025persona, zhou2026emoshift, facchiano2025activation, kang2025model, sammani2026negation}, we adopt a training-free steering method to mitigate negation bias in audio-language retrieval models.
Similar to observations in vision-language embeddings~\cite{seeing_negation_2026}, we find an affirmation bias, where captions such as ``dog barking'' and ``no dog barking'' are mapped to highly similar embeddings, leading to confusion between presence and absence.
To counter this, we edit the caption embedding directly. 
Given a caption $C$ containing a negated concept $C_{\text{neg}}$, we compute
$e^{*} = e_{C} - \lambda e_{C_{\text{neg}}}$
, where $e_{C}$ is the full caption embedding and $e_{C_{\text{neg}}}$ is the embedding of the negated sound concepts (e.g., ``dog barking''). 
We set $\lambda=0.2$ for all models, datasets, and tasks, rather than tuning it per setting. 
This provides a fair, non-oracle comparison and keeps the intervention in a small-perturbation regime, where the embedding is steered away from the negated concept while preserving the remaining caption semantics. 
We report full $\lambda$-sweep results in Appendix~\ref{steering-lamda-sweep}.

\begin{table*}[htbp]
\centering
\caption{
Text-to-audio retrieval performance on standard and negation-aware queries under AudioCaps and Clotho.
For each column, the \textbf{best} and \underline{second-best} scores are highlighted in bold and underline, respectively.
}
\small
\begin{tabular}{l ccc ccc ccc ccc}
\toprule
& \multicolumn{6}{c}{\textbf{AudioCaps test~\cite{kim2019audiocaps}}} & \multicolumn{6}{c}{\textbf{Clotho evaluation~\cite{drossos2020clotho}}} \\
\cmidrule(lr){2-7} \cmidrule(lr){8-13}
\textbf{Model} 
& \multicolumn{3}{c}{Standard} 
& \multicolumn{3}{c}{Retrieval-Neg}
& \multicolumn{3}{c}{Standard} 
& \multicolumn{3}{c}{Retrieval-Neg} \\
\cmidrule(lr){2-4} \cmidrule(lr){5-7} \cmidrule(lr){8-10} \cmidrule(lr){11-13}
& R@1 & R@5 & R@10 & R@1 & R@5 & R@10
& R@1 & R@5 & R@10 & R@1 & R@5 & R@10 \\
\midrule
\textit{LAION-CLAP}~\cite{wu2023large} \\
-- 630k-audioset-best 
& 35.3 & 70.1 & 82.8 
& 26.5 & 60.3 & 75.6
& 14.8 & 37.3 & 51.2 
& 14.4 & 35.4 & 48.7 \\
-- 630k-best
& 31.4 & 67.5 & 80.8 
& 26.3 & 59.1 & 75.8
& 15.3 & 37.4 & 51.6 
& 16.1 & 37.0 & 50.6 \\
-- music-audioset
& 33.0 & \underline{71.9} & 85.0 
& 23.5 & 57.0 & 74.8
& 13.9 & 35.2 & 49.2 
& 13.1 & 33.8 & 46.1 \\
-- music-speech-audioset
& 35.0 & 68.6 & 84.9 
& 21.7 & 57.9 & 73.4
& 14.4 & 37.7 & 50.9 
& 12.2 & 34.2 & 46.8 \\
-- music-speech
& 30.6 & 67.0 & 80.5 
& 24.5 & 59.4 & 75.6
& 14.9 & 38.2 & 51.2 
& 14.8 & \underline{38.0} & 50.1 \\
\textit{M2D-CLAP}~\cite{niizumi2025m2d-clap}
\\
-- M2D-CLAP-2025
& \underline{40.9} & \textbf{77.9} & \textbf{89.2}
& \underline{29.9} & \underline{63.2} & \underline{77.1}
& \underline{18.9} & \underline{43.9} & \underline{58.1}
& \underline{16.2} & \underline{38.0} & \underline{51.4}
\\
-- M2D-CLAP-2024
& 22.6 & 55.0 & 70.6
& 16.9 & 46.1 & 62.1
& 9.6 & 24.5 & 32.6
& 6.5 & 19.6 & 27.4
\\
\textit{WAVE}~\cite{tang2025wave} \\
-- WAVE-7B
& \textbf{44.0} & \textbf{77.9} & \underline{88.7}
& \textbf{39.6} & \textbf{75.2} & \textbf{86.1}
& \textbf{26.6} & \textbf{52.7} & \textbf{66.3}
& \textbf{22.1} & \textbf{50.0} & \textbf{61.5}
\\
\bottomrule
\end{tabular}
\label{tab:retrieval-neg}
\end{table*}
\section{Experimental Setup}

\textbf{Evaluation data construction.}
We construct NegEval-Audio on the test split of AudioCaps~\cite{kim2019audiocaps} and the evaluation split of Clotho~\cite{drossos2020clotho}.
For LLM-based components, we use Claude Opus 4.8 for text generation, including caption refinement and MCQ construction.
For audio-based sound-existence verification, we use the audio-aware LLM \textit{Qwen2.5-Omni-7B}~\cite{xu2025qwen2}.
We choose this model due to its strong performance on recent audio-related benchmarks~\cite{huang2024dynamic, huang2024dynamic2, kuan2024understanding, kuan2024can, sakshi2024mmau, ma2025mmar, lu2025speech, wang2025mmsu, yang2025paras2s, chang2025game, kuan2026, kuan2026aqascore, kuan2026walking}.

\textbf{Evaluated models.}
We evaluate a broad set of publicly available audio--text embedding models. We first consider LAION-CLAP checkpoints~\cite{wu2023large}, covering general audio, music, speech, and their combinations. 
We then include M2D-CLAP~\cite{niizumi2025m2d-clap, niizumi2024m2d-clap}, which departs from the original CLAP design through architectural and training modifications. Finally, we evaluate WAVE~\cite{tang2025wave}, a recent multimodal embedding model built upon multimodal LLMs, to reflect the emerging trend of MLLM-based audio--text representation learning.

\textbf{Evaluation protocol.}
To isolate the effect of negation, we compare standard Retrieval with Retrieval-Neg and evaluate MCQ-Neg.
In text-to-audio retrieval, queries rank candidate audio samples by embedding similarity, and performance is measured by Recall@K.
We report K equal to 1, 5, and 10.
The retrieval protocol is identical across settings; only the query semantics differ.
For MCQ-Neg, each question contains four candidate descriptions with one correct answer.
The model selects the option whose text embedding has the highest cosine similarity to the audio embedding; no generation or task-specific tuning is involved. 
We report accuracy for each question type and their overall average to analyze model behavior under affirmation, negation, and hybrid conditions.

\begin{table*}[t]
\centering
\caption{
MCQ-Neg performance (\%) across three question types on AudioCaps and Clotho.
\textbf{Best} and \underline{second-best} scores are highlighted.
}
\small
\setlength{\tabcolsep}{5.5pt}
\begin{tabular}{l cccc cccc}
\toprule
& \multicolumn{4}{c}{\textbf{AudioCaps test~\cite{kim2019audiocaps}}} 
& \multicolumn{4}{c}{\textbf{Clotho evaluation~\cite{drossos2020clotho}}} \\
\cmidrule(lr){2-5} \cmidrule(lr){6-9}
\textbf{Model} 
& Affirmation & Negation & Hybrid & Average
& Affirmation & Negation & Hybrid & Average \\
\midrule
{\textit{LAION-CLAP}}~\cite{wu2023large} \\
-- 630k-audioset-best 
& \underline{81.1} & 0.6 & \underline{47.8} & 43.2
& 59.8 & 1.9 & \textbf{49.9} & 37.2 \\
-- 630k-best
& 80.9 & 0.6 & 44.1 & 41.9
& \underline{66.5} & 3.5 & 44.2 & 38.1 \\
-- music-audioset
& \textbf{85.4} & 0.6 & 43.8 & \underline{43.3}
& 59.6 & \textbf{7.1} & 48.3 & \underline{38.4} \\
-- music-speech-audioset
& 80.8 & 0.3 & 42.1 & 41.1
& 56.1 & 3.9 & 32.8 & 30.9 \\
-- music-speech
& 74.1 & \textbf{1.1} & 29.6 & 34.9
& 49.0 & 3.5 & 31.3 & 27.9 \\
{\textit{M2D-CLAP}}~\cite{niizumi2025m2d-clap} \\
-- M2D-CLAP-2025
& 69.4 & 0.2 & 14.0 & 27.9
& 42.7 & 3.9 & 19.7 & 22.1 
\\
-- M2D-CLAP-2024
& 67.4 & 0.6 & 24.2 & 30.8
& 41.1 & 4.9 & 23.7 & 23.2
\\
{\textit{WAVE}}~\cite{tang2025wave} \\
-- WAVE-7B
& 78.3 & \underline{0.9} & \textbf{51.3} & \textbf{43.5}
& \textbf{75.0} & \underline{5.3} & \underline{49.2} & \textbf{42.2} \\
\bottomrule
\end{tabular}
\label{tab:mcq-neg}
\end{table*}

\begin{table*}[htbp]
\centering
\caption{
Effect of steering ($\lambda{=}0.2$) on negation-aware tasks:
Retrieval-Neg and MCQ-Neg. 
The \textbf{best} and \underline{second-best} results in each column are shown in bold and underlined, respectively. 
$\Delta$ denotes the change after steering relative to the corresponding baseline; $\Delta$ columns are shaded light blue.
}
\small
\begin{tabular}{l ccc ccc ccc ccc}
\toprule
& \multicolumn{6}{c}{\textbf{AudioCaps test~\cite{kim2019audiocaps}}} & \multicolumn{6}{c}{\textbf{Clotho evaluation~\cite{drossos2020clotho}}} \\
\cmidrule(lr){2-7} \cmidrule(lr){8-13}
\textbf{Model}
& \multicolumn{3}{c}{Retrieval-Neg (R@5)}
& \multicolumn{3}{c}{MCQ-Neg (Avg)}
& \multicolumn{3}{c}{Retrieval-Neg (R@5)}
& \multicolumn{3}{c}{MCQ-Neg (Avg)} \\
\cmidrule(lr){2-4} \cmidrule(lr){5-7} \cmidrule(lr){8-10} \cmidrule(lr){11-13}
& Before & After & \cellcolor{lightblue}$\Delta$
& Before & After & \cellcolor{lightblue}$\Delta$
& Before & After & \cellcolor{lightblue}$\Delta$
& Before & After & \cellcolor{lightblue}$\Delta$ \\
\midrule
\textit{LAION-CLAP}~\cite{wu2023large} \\
-- 630k-audioset-best
& 60.3 & 61.8 & \cellcolor{lightblue}+1.5
& 43.2 & \underline{45.0} & \cellcolor{lightblue}+1.8
& 35.4 & 36.7 & \cellcolor{lightblue}+1.3
& 37.2 & 36.6 & \cellcolor{lightblue}-0.6 \\
-- 630k-best
& 59.1 & 58.5 & \cellcolor{lightblue}-0.6
& 41.9 & 42.8 & \cellcolor{lightblue}+0.9
& 37.0 & 38.7 & \cellcolor{lightblue}+1.7
& 38.1 & \underline{39.4} & \cellcolor{lightblue}+1.3 \\
-- music-audioset
& 57.0 & 57.3 & \cellcolor{lightblue}+0.3
& \underline{43.3} & 44.9 & \cellcolor{lightblue}+1.6
& 33.8 & 35.0 & \cellcolor{lightblue}+1.2
& \underline{38.4} & 37.7 & \cellcolor{lightblue}-0.7 \\
-- music-speech-audioset
& 57.9 & 58.6 & \cellcolor{lightblue}+0.7
& 41.1 & 42.2 & \cellcolor{lightblue}+1.1
& 34.2 & 34.8 & \cellcolor{lightblue}+0.6
& 30.9 & 31.1 & \cellcolor{lightblue}+0.2 \\
-- music-speech
& 59.4 & 58.1 & \cellcolor{lightblue}-1.3
& 34.9 & 38.4 & \cellcolor{lightblue}\underline{+3.5}
& \underline{38.0} & 37.5 & \cellcolor{lightblue}-0.5
& 27.9 & 29.5 & \cellcolor{lightblue}\underline{+1.6} \\
\textit{M2D-CLAP}~\cite{niizumi2025m2d-clap} \\
-- M2D-CLAP-2025
& \underline{63.2} & \underline{65.8} & \cellcolor{lightblue}\underline{+2.6}
& 27.9 & 30.2 & \cellcolor{lightblue}+2.3
& \underline{38.0} & \underline{42.3} & \cellcolor{lightblue}\textbf{+4.3}
& 22.1 & 22.9 & \cellcolor{lightblue}+0.8 \\
-- M2D-CLAP-2024
& 46.1 & 50.2 & \cellcolor{lightblue}\textbf{+4.1}
& 30.8 & 30.9 & \cellcolor{lightblue}+0.1
& 19.6 & 22.5 & \cellcolor{lightblue}\underline{+2.9}
& 23.2 & 24.1 & \cellcolor{lightblue}+0.9 \\
\textit{WAVE}~\cite{tang2025wave} \\
-- WAVE-7B
& \textbf{75.2} & \textbf{77.1} & \cellcolor{lightblue}+1.9
& \textbf{43.5} & \textbf{54.5} & \cellcolor{lightblue}\textbf{+11.0}
& \textbf{50.0} & \textbf{52.0} & \cellcolor{lightblue}+2.0
& \textbf{42.2} & \textbf{52.4} & \cellcolor{lightblue}\textbf{+10.2} \\
\bottomrule
\end{tabular}
\label{tab:steering}
\end{table*}

\section{Results and Analysis}

\subsection{Standard Retrieval Masks Negation Failure}

\textbf{Standard retrieval looks reliable, but negation-aware retrieval exposes a hidden failure.}
Table~\ref{tab:retrieval-neg} compares standard text-to-audio retrieval with Retrieval-Neg, where the audio candidates and evaluation protocol are unchanged, and only the query is modified to include an explicit absent sound concept. 
On AudioCaps, all models degrade under Retrieval-Neg. Averaged over the eight checkpoints, R@1 drops from 34.1 to 26.1, R@5 from 69.5 to 59.8, and R@10 from 82.8 to 75.1. 
The drop is also clear for strong models: M2D-CLAP-2025 falls from 40.9 to 29.9 in R@1, and LAION-CLAP music-speech-audioset from 35.0 to 21.7. 
On Clotho, the degradation is smaller but consistent on average, with R@1, R@5, and R@10 dropping from 16.1, 38.4, and 51.4 to 14.4, 35.8, and 47.8, respectively. 
These results show that affirmation-only retrieval can overestimate semantic understanding: retrieving present sounds well does not imply that a model can use negated sound concepts as exclusion constraints.

\subsection{MCQ-Neg Reveals an Affirmation Bias}

\textbf{MCQ-Neg confirms that the failure is specific to negation, not general audio recognition.}
Table~\ref{tab:mcq-neg} breaks performance into Affirmation, Negation, and Hybrid questions. 
For Affirmation questions, where the correct option describes sounds present in the audio, models remain substantially above chance: accuracy ranges from 67.4\% to 85.4\% on AudioCaps and from 41.1\% to 75.0\% on Clotho. 
In contrast, Negation accuracy collapses to 0.2--1.1\% on AudioCaps and 1.9--7.1\% on Clotho, far below the 25\% random baseline for four-way multiple choice. 
This near-zero performance indicates a systematic affirmation bias: mentioned sound concepts are treated as positive evidence even when explicitly negated. 
Hybrid questions show an intermediate pattern, with accuracy ranging from 14.0\% to 51.3\% on AudioCaps and from 19.7\% to 49.9\% on Clotho. 
Together, these results show that current audio--language embeddings can recognize present events, but fail to bind negation to the intended sound concept.

\subsection{Steering Helps Local Decisions but Not Global Retrieval}

\textbf{Steering shows that the failure is partially editable, but not fully solved.}
Table~\ref{tab:steering} reports the effect of training-free steering with $\lambda=0.2$. 
For Retrieval-Neg, gains are limited: R@5 changes by +1.2 points on average on AudioCaps, with per-model changes from -1.3 to +4.1, and by +1.7 points on Clotho, with changes from -0.5 to +4.3. 
This suggests that steering slightly reduces the attraction toward negated concepts, but does not substantially reshape the global retrieval space. 
The effect is clearer for MCQ-Neg, especially on WAVE-7B, which improves from 43.5\% to 54.5\% on AudioCaps and from 42.2\% to 52.4\% on Clotho. 
For CLAP and M2D-CLAP checkpoints, however, MCQ gains are modest, ranging from +0.1 to +3.5 points on AudioCaps and from -0.7 to +1.6 points on Clotho. 
This contrast suggests that steering can alter local option rankings in MCQ-Neg, but full-corpus retrieval requires a more globally consistent geometry where audio containing the forbidden sound is pushed away. 
The limited retrieval gains therefore suggest that negation failure is not a surface artifact of the word ``no'', but reflects a deeper weakness in the joint embedding space.
Details on the selection of $\lambda$ are provided in Appendix~\ref{steering-lamda-sweep}.

\begin{table}[t]
\centering
\small
\caption{
Controlled diagnosis of CLAP text-side negation handling. 
Values are averaged over five LAION-CLAP checkpoints. 
}
\label{tab:controlled_diagnosis}
\begin{tabular}{l c}
\toprule
\textbf{Diagnostic} & \textbf{Mean} \\
\midrule
\multicolumn{2}{l}{\textit{Polarity sensitivity}} \\
Normalized affinity to affirmative text & 1.07 \\
Polarity change / content swap & 0.30 \\
\midrule
\multicolumn{2}{l}{\textit{Negation scope}} \\
Scope permutation similarity & 0.70 \\
Content swap similarity & 0.58 \\
Bag-of-concepts similarity & 0.82 \\
\midrule
\multicolumn{2}{l}{\textit{Audio retrieval bridge}} \\
Corr. between negated and affirmative margins & 0.76 \\
Negated-query margin & +0.018 \\
Affirmative-query margin & +0.072 \\
\bottomrule
\end{tabular}
\end{table}

\subsection{Text-Side Diagnosis of Negation Failure}
\label{sec:textdiag}

The retrieval results reveal a recurring failure mode in CLAP-style audio--text retrieval: negated queries are often matched to audio containing the very sound that should be excluded. 
To understand this mechanism, we conduct a controlled diagnosis on five LAION-CLAP checkpoints. 
Our goal is not to introduce another benchmark, but to isolate whether a negated description, such as ``$P$, with no $Y$'', is sufficiently separated from its affirmative counterpart, ``$P$, with $Y$'', in the embedding space.
To eliminate confounding variables in real-world datasets, such as complex free-form captions and noisy acoustic backgrounds, we construct a clean and controlled diagnostic set using ESC-50~\cite{piczak2015dataset}. 
This design allows us to isolate the role of negation more precisely.
Each item contains two positive sound events and one designated negated concept $Y$. We render matched target--distractor audio pairs that share the same base mixture and differ only in whether $Y$ is present: the distractor contains $Y$, while the target replaces it with a matched filler sound from the same category group and with similar loudness. This gives 1,000 controlled items where the presence or absence of each sound event is known by construction, avoiding the verification noise of caption-derived negatives. Table~\ref{tab:controlled_diagnosis} summarizes the diagnostic results averaged over the five checkpoints.

\textbf{Negation barely moves the text embedding.}
We first compare each negated query with its affirmative counterpart. Since raw cosine similarities in CLAP text space are difficult to interpret due to anisotropy, we calibrate them using within-query references: an unrelated query as the floor and a meaning-preserving paraphrase as the ceiling. We report normalized affinity, where 0 corresponds to the unrelated floor and 1 corresponds to the paraphrase ceiling. Under this calibration, the affirmative counterpart reaches the paraphrase ceiling, with a normalized affinity of 1.07. In other words, ``$P$, with $Y$'' is as close to ``$P$, with no $Y$'' as a meaning-preserving paraphrase is. We further compare this polarity change against a content-swap reference, where one mentioned sound concept is replaced. The polarity change produces only 30\% of the embedding distance caused by the content swap, suggesting that the text encoder treats negation as a weak perturbation rather than a distinct semantic condition.

\textbf{Negation is weakly bound to the intended concept.}
We then test whether the encoder tracks which sound concept is negated. For each query, we construct three variants that mention the same concepts but negate a different one. If negation scope were well represented, these variants should be clearly separated. To make this comparison interpretable, we again use content swap as a reference scale: it measures how much the embedding changes when one mentioned sound concept is replaced. As shown in Table~\ref{tab:controlled_diagnosis}, scope variants remain more similar to each other than content-swapped variants, and the query is also highly similar to its bag-of-concepts form, which removes function and polarity words and keeps only the mentioned sound concepts. These results indicate that the encoder preserves which concepts are mentioned more strongly than whether a specific concept is negated.

\textbf{The text-side collapse propagates to retrieval.}
Finally, we examine whether this text-side behavior is reflected in audio retrieval. For each item, we compute the audio margin between the $Y$-containing distractor and the $Y$-absent target:
$m(q)=s(q,a_{\mathrm{dist}})-s(q,a_{\mathrm{tgt}})$,
where $a_{\mathrm{dist}}$ and $a_{\mathrm{tgt}}$ denote the distractor and target audio, respectively. A correctly interpreted negated query should yield a negative margin, preferring the target. Instead, the average margin for negated queries remains positive, meaning that they still prefer audio containing the forbidden sound. The affirmative query yields a stronger positive margin, and the two margins are strongly correlated across items. This means that examples where the affirmative query strongly prefers the $Y$-containing audio are also examples where the negated query tends to prefer it. Thus, negated queries behave like attenuated affirmative queries rather than reversing the preference toward the $Y$-absent audio.
More details are provided in Appendix~\ref{app:controlled}.
Together, these diagnostics suggest that CLAP under-encodes negation in its text space. A negated description remains close to a bag of mentioned sound concepts and to its affirmative counterpart, so the joint embedding space has little basis for retrieving audio that excludes the negated sound.

\section{Conclusion}

We presented NegEval-Audio, a framework for evaluating negation understanding in audio-language embedding models. 
Our results reveal a systematic affirmation bias: models can match sounds that are present, but often fail to treat negated sounds as constraints on what should be absent. 
To address this, future work should explore explicit negation-aware training objectives~\cite{xu2026omni} and compositional alignment. 
Ultimately, these findings call for audio-language models that go beyond what is heard and account for what should not be heard.

\section{Limitations}
First, our steering method assumes access to the negated concept. 
While this concept is known by construction in NegEval-Audio, applying the method to free-form queries would require an additional parsing step to identify the negated span. 
Second, each Retrieval-Neg query contains only a single negated concept; queries involving multiple, coordinated, or nested negations therefore remain unexplored. 
Third, NegEval-Audio is constructed from AudioCaps and Clotho, both of which primarily cover general environmental sounds. 
Extending the benchmark to other domains, such as spoken content and musical attributes, is an important direction for future work. 
Finally, we use a fixed steering strength, $\lambda$, across all models to ensure a fair and controlled comparison. 
Although this avoids model-specific tuning, selecting $\lambda$ separately for each model may produce different trade-offs between negation sensitivity and overall retrieval performance.

\section*{Generative AI Use Disclosure}
The authors used ChatGPT and Claude to assist with grammar checking, language polishing, and improving the readability of the manuscript. 

\section*{Acknowledgments}
We thank Kai-Wei Chang for insightful discussions and valuable feedback, which helped improve this work.


\bibliographystyle{IEEEtran}
\bibliography{refs}

@inproceedings{elizalde2023clap,
  title={Clap learning audio concepts from natural language supervision},
  author={Elizalde, Benjamin and Deshmukh, Soham and Al Ismail, Mahmoud and Wang, Huaming},
  booktitle={ICASSP 2023-2023 IEEE International Conference on Acoustics, Speech and Signal Processing (ICASSP)},
  pages={1--5},
  year={2023},
  organization={IEEE}
}

@inproceedings{wu2023large,
  title={Large-scale contrastive language-audio pretraining with feature fusion and keyword-to-caption augmentation},
  author={Wu, Yusong and Chen, Ke and Zhang, Tianyu and Hui, Yuchen and Berg-Kirkpatrick, Taylor and Dubnov, Shlomo},
  booktitle={ICASSP 2023-2023 IEEE International Conference on Acoustics, Speech and Signal Processing (ICASSP)},
  pages={1--5},
  year={2023},
  organization={IEEE}
}

@article{niizumi2025m2d-clap,
    author={Niizumi, Daisuke and Takeuchi, Daiki and Yasuda, Masahiro and Thien Nguyen, Binh and Ohishi, Yasunori and Harada, Noboru},
    journal={IEEE Access}, 
    title={M2D-CLAP: Exploring General-Purpose Audio-Language Representations Beyond CLAP}, 
    year={2025},
    volume={13},
    number={},
    pages={163313-163330},
    doi={10.1109/ACCESS.2025.3611348}}

@inproceedings{niizumi2024m2d-clap,
    title   = {{M2D-CLAP: Masked Modeling Duo Meets CLAP for Learning General-purpose Audio-Language Representation}},
    author  = {Daisuke Niizumi and Daiki Takeuchi and Yasunori Ohishi and Noboru Harada and Masahiro Yasuda and Shunsuke Tsubaki and Keisuke Imoto},
    booktitle={Interspeech},
    year    = {2024},
    pages   = {57--61},
    doi     = {10.21437/Interspeech.2024-29}}

@inproceedings{yuan2024t,
  title={T-clap: Temporal-enhanced contrastive language-audio pretraining},
  author={Yuan, Yi and Chen, Zhuo and Liu, Xubo and Liu, Haohe and Xu, Xuenan and Jia, Dongya and Chen, Yuanzhe and Plumbley, Mark D and Wang, Wenwu},
  booktitle={2024 IEEE 34th International Workshop on Machine Learning for Signal Processing (MLSP)},
  pages={1--6},
  year={2024},
  organization={IEEE}
}

@inproceedings{li2024advancing,
  title={Advancing multi-grained alignment for contrastive language-audio pre-training},
  author={Li, Yiming and Guo, Zhifang and Wang, Xiangdong and Liu, Hong},
  booktitle={Proceedings of the 32nd ACM International Conference on Multimedia},
  pages={7356--7365},
  year={2024}
}

@inproceedings{kim2019audiocaps,
  title={Audiocaps: Generating captions for audios in the wild},
  author={Kim, Chris Dongjoo and Kim, Byeongchang and Lee, Hyunmin and Kim, Gunhee},
  booktitle={Proceedings of the 2019 Conference of the North American Chapter of the Association for Computational Linguistics: Human Language Technologies, Volume 1 (Long and Short Papers)},
  pages={119--132},
  year={2019}
}

@inproceedings{drossos2020clotho,
  title={Clotho: An audio captioning dataset},
  author={Drossos, Konstantinos and Lipping, Samuel and Virtanen, Tuomas},
  booktitle={ICASSP 2020-2020 IEEE International Conference on Acoustics, Speech and Signal Processing (ICASSP)},
  pages={736--740},
  year={2020},
  organization={IEEE}
}

@inproceedings{ghoshcompa,
  title={CompA: Addressing the Gap in Compositional Reasoning in Audio-Language Models},
  author={Ghosh, Sreyan and Seth, Ashish and Kumar, Sonal and Tyagi, Utkarsh and Evuru, Chandra Kiran Reddy and Sakshi, S and Nieto, Oriol and Duraiswami, Ramani and Manocha, Dinesh and others},
  booktitle={The Twelfth International Conference on Learning Representations},
  year={2024}
}

@inproceedings{piczak2015dataset,
  title = {{ESC}: {Dataset} for {Environmental Sound Classification}},
  author = {Piczak, Karol J.},
  booktitle = {Proceedings of the 23rd {Annual ACM Conference} on {Multimedia}},
  publisher = {{ACM Press}},
  pages = {1015--1018},
  year = {2015}
}

@article{chu2023qwen,
  title={Qwen-audio: Advancing universal audio understanding via unified large-scale audio-language models},
  author={Chu, Yunfei and Xu, Jin and Zhou, Xiaohuan and Yang, Qian and Zhang, Shiliang and Yan, Zhijie and Zhou, Chang and Zhou, Jingren},
  journal={arXiv preprint arXiv:2311.07919},
  year={2023}
}

@article{chu2024qwen2,
  title={Qwen2-audio technical report},
  author={Chu, Yunfei and Xu, Jin and Yang, Qian and Wei, Haojie and Wei, Xipin and Guo, Zhifang and Leng, Yichong and Lv, Yuanjun and He, Jinzheng and Lin, Junyang and others},
  journal={arXiv preprint arXiv:2407.10759},
  year={2024}
}

@inproceedings{tang2023salmonn,
  title={SALMONN: Towards Generic Hearing Abilities for Large Language Models},
  author={Tang, Changli and Yu, Wenyi and Sun, Guangzhi and Chen, Xianzhao and Tan, Tian and Li, Wei and Lu, Lu and MA, Zejun and Zhang, Chao},
  booktitle={The Twelfth International Conference on Learning Representations},
  year={2024}
}

@inproceedings{gong2023joint,
  title={Joint audio and speech understanding},
  author={Gong, Yuan and Liu, Alexander H and Luo, Hongyin and Karlinsky, Leonid and Glass, James},
  booktitle={2023 IEEE Automatic Speech Recognition and Understanding Workshop (ASRU)},
  pages={1--8},
  year={2023},
  organization={IEEE}
}

@inproceedings{gong2023listen,
  title={Listen, Think, and Understand},
  author={Gong, Yuan and Luo, Hongyin and Liu, Alexander H and Karlinsky, Leonid and Glass, James R},
  booktitle={The Twelfth International Conference on Learning Representations},
  year={2024}
}

@inproceedings{kuan2024speech,
  title={Speech-copilot: Leveraging large language models for speech processing via task decomposition, modularization, and program generation},
  author={Kuan, Chun-Yi and Yang, Chih-Kai and Huang, Wei-Ping and Lu, Ke-Han and Lee, Hung-yi},
  booktitle={2024 IEEE Spoken Language Technology Workshop (SLT)},
  pages={1060--1067},
  year={2024},
  organization={IEEE}
}

@inproceedings{fathullah2023towards,
  title={Audiochatllama: Towards general-purpose speech abilities for llms},
  author={Fathullah, Yassir and Wu, Chunyang and Lakomkin, Egor and Li, Ke and Jia, Junteng and Shangguan, Yuan and Mahadeokar, Jay and Kalinli, Ozlem and Fuegen, Christian and Seltzer, Mike},
  booktitle={Proceedings of the 2024 Conference of the North American Chapter of the Association for Computational Linguistics: Human Language Technologies (Volume 1: Long Papers)},
  pages={5522--5532},
  year={2024}
}

@article{wang2023blsp,
  title={Blsp: Bootstrapping language-speech pre-training via behavior alignment of continuation writing},
  author={Wang, Chen and Liao, Minpeng and Huang, Zhongqiang and Lu, Jinliang and Wu, Junhong and Liu, Yuchen and Zong, Chengqing and Zhang, Jiajun},
  journal={arXiv preprint arXiv:2309.00916},
  year={2023}
}

@inproceedings{wang2024blsp,
  title={BLSP-Emo: Towards Empathetic Large Speech-Language Models},
  author={Wang, Chen and Liao, Minpeng and Huang, Zhongqiang and Wu, Junhong and Zong, Chengqing and Zhang, Jiajun},
  booktitle={Proceedings of the 2024 Conference on Empirical Methods in Natural Language Processing},
  pages={19186--19199},
  year={2024}
}

@inproceedings{ghoshaudio,
  title={Audio Flamingo 2: An Audio-Language Model with Long-Audio Understanding and Expert Reasoning Abilities},
  author={Ghosh, Sreyan and Kong, Zhifeng and Kumar, Sonal and Sakshi, S and Kim, Jaehyeon and Ping, Wei and Valle, Rafael and Manocha, Dinesh and Catanzaro, Bryan},
  booktitle={Forty-second International Conference on Machine Learning},
  year={2025}
}

@inproceedings{goel2025audio,
  title={Audio Flamingo 3: Advancing Audio Intelligence with Fully Open Large Audio Language Models},
  author={Ghosh, Sreyan and Goel, Arushi and Kim, Jaehyeon and Kumar, Sonal and Kong, Zhifeng and Lee, Sang-gil and Yang, Chao-Han Huck and Duraiswami, Ramani and Manocha, Dinesh and Valle, Rafael and others},
  booktitle={The Thirty-ninth Annual Conference on Neural Information Processing Systems},
  year={2025}
}

@article{abouelenin2025phi,
  title={Phi-4-mini technical report: Compact yet powerful multimodal language models via mixture-of-loras},
  author={Abouelenin, Abdelrahman and Ashfaq, Atabak and Atkinson, Adam and Awadalla, Hany and Bach, Nguyen and Bao, Jianmin and Benhaim, Alon and Cai, Martin and Chaudhary, Vishrav and Chen, Congcong and others},
  journal={arXiv preprint arXiv:2503.01743},
  year={2025}
}

@article{xu2025qwen2,
  title={Qwen2.5-omni technical report},
  author={Xu, Jin and Guo, Zhifang and He, Jinzheng and Hu, Hangrui and He, Ting and Bai, Shuai and Chen, Keqin and Wang, Jialin and Fan, Yang and Dang, Kai and others},
  journal={arXiv preprint arXiv:2503.20215},
  year={2025}
}

@article{comanici2025gemini,
  title={Gemini 2.5: Pushing the frontier with advanced reasoning, multimodality, long context, and next generation agentic capabilities},
  author={Comanici, Gheorghe and Bieber, Eric and Schaekermann, Mike and Pasupat, Ice and Sachdeva, Noveen and Dhillon, Inderjit and Blistein, Marcel and Ram, Ori and Zhang, Dan and Rosen, Evan and others},
  journal={arXiv preprint arXiv:2507.06261},
  year={2025}
}

@article{achiam2023gpt,
  title={Gpt-4 technical report},
  author={Achiam, Josh and Adler, Steven and Agarwal, Sandhini and Ahmad, Lama and Akkaya, Ilge and Aleman, Florencia Leoni and Almeida, Diogo and Altenschmidt, Janko and Altman, Sam and Anadkat, Shyamal and others},
  journal={arXiv preprint arXiv:2303.08774},
  year={2023}
}

@article{kuan2025alignment,
  title   = {From Alignment to Advancement: Bootstrapping Audio-Language Alignment With Synthetic Data},
  author  = {Kuan, Chun-Yi and Lee, Hung-yi},
  journal = {IEEE Transactions on Audio, Speech and Language Processing},
  year    = {2025},
  volume  = {33},
  pages   = {4604--4619},
  doi     = {10.1109/TASLPRO.2025.3626233}
}

@article{yang2024building,
  title={Building a taiwanese mandarin spoken language model: A first attempt},
  author={Yang, Chih-Kai and Fu, Yu-Kuan and Li, Chen-An and Lin, Yi-Cheng and Lin, Yu-Xiang and Chen, Wei-Chih and Chung, Ho Lam and Kuan, Chun-Yi and Huang, Wei-Ping and Lu, Ke-Han and others},
  journal={arXiv preprint arXiv:2411.07111},
  year={2024}
}

@inproceedings{kuan2025teaching,
  title={Teaching Audio-Aware Large Language Models What Does Not Hear: Mitigating Hallucinations through Synthesized Negative Samples},
  author={Kuan, Chun-Yi and Lee, Hung-yi},
  booktitle={Proc. Interspeech 2025},
  pages={2073--2077},
  year={2025}
}

@inproceedings{held2025distilling,
  title={Distilling an end-to-end voice assistant without instruction training data},
  author={Held, William and Zhang, Yanzhe and Li, Minzhi and Shi, Weiyan and Ryan, Michael J and Yang, Diyi},
  booktitle={Proceedings of the 63rd Annual Meeting of the Association for Computational Linguistics (Volume 1: Long Papers)},
  pages={7876--7891},
  year={2025}
}

@inproceedings{chang2025game,
  title={Game-Time: Evaluating Temporal Dynamics in Spoken Language Models},
  author={Chang, Kai-Wei and Hu, En-Pei and Kuan, Chun-Yi and Ren, Wenze and Chen, Wei-Chih and Lin, Guan-Ting and Tsao, Yu and Sun, Shao-Hua and Lee, Hung-yi and Glass, James},
  booktitle={ICASSP 2026-2026 IEEE International Conference on Acoustics, Speech and Signal Processing (ICASSP)},
  pages={1--5},
  year={2026},
}

@article{ma2025mmar,
  title={MMAR: A Challenging Benchmark for Deep Reasoning in Speech, Audio, Music, and Their Mix},
  author={Ma, Ziyang and Ma, Yinghao and Zhu, Yanqiao and Yang, Chen and Chao, Yi-Wen and Xu, Ruiyang and others},
  journal={arXiv preprint arXiv:2505.13032},
  year={2025}
}

@article{wang2025mmsu,
  title={Mmsu: A massive multi-task spoken language understanding and reasoning benchmark},
  author={Wang, Dingdong and Wu, Jincenzi and Li, Junan and Yang, Dongchao and Chen, Xueyuan and Zhang, Tianhua and Meng, Helen},
  journal={arXiv preprint arXiv:2506.04779},
  year={2025}
}

@inproceedings{huang2024dynamic,
  title={Dynamic-superb: Towards a dynamic, collaborative, and comprehensive instruction-tuning benchmark for speech},
  author={Huang, Chien-yu and Lu, Ke-Han and Wang, Shih-Heng and Hsiao, Chi-Yuan and Kuan, Chun-Yi and Wu, Haibin and Arora, Siddhant and Chang, Kai-Wei and Shi, Jiatong and Peng, Yifan and others},
  booktitle={ICASSP 2024-2024 IEEE International Conference on Acoustics, Speech and Signal Processing (ICASSP)},
  pages={12136--12140},
  year={2024},
  organization={IEEE}
}

@inproceedings{huang2024dynamic2,
  title={Dynamic-SUPERB Phase-2: A Collaboratively Expanding Benchmark for Measuring the Capabilities of Spoken Language Models with 180 Tasks},
  author={Huang, Chien-yu and Chen, Wei-Chih and Yang, Shu-wen and Liu, Andy T and Li, Chen-An and Lin, Yu-Xiang and Tseng, Wei-Cheng and Diwan, Anuj and Shih, Yi-Jen and Shi, Jiatong and others},
  booktitle={The Thirteenth International Conference on Learning Representations},
  year={2025}
}

@inproceedings{kuan2024can,
  title={Can large audio-language models truly hear? tackling hallucinations with multi-task assessment and stepwise audio reasoning},
  author={Kuan, Chun-Yi and Lee, Hung-yi},
  booktitle={ICASSP 2025-2025 IEEE International Conference on Acoustics, Speech and Signal Processing (ICASSP)},
  pages={1--5},
  year={2025},
  organization={IEEE}
}

@inproceedings{kuan2024understanding,
  title={Understanding Sounds, Missing the Questions: The Challenge of Object Hallucination in Large Audio-Language Models},
  author={Kuan, Chun-Yi and Huang, Wei-Ping and Lee, Hung-yi},
  booktitle={Interspeech 2024},
  year={2024}
}

@inproceedings{lu2025speech,
  title={Speech-IFEval: Evaluating Instruction-Following and Quantifying Catastrophic Forgetting in Speech-Aware Language Models},
  author={Lu, Ke-Han and Kuan, Chun-Yi and Lee, Hung-yi},
  booktitle={Proc. Interspeech 2025},
  pages={2078--2082},
  year={2025}
}

@article{chang2024speechprompt,
  title={Speechprompt: Prompting speech language models for speech processing tasks},
  author={Chang, Kai-Wei and Wu, Haibin and Wang, Yu-Kai and Wu, Yuan-Kuei and Shen, Hua and Tseng, Wei-Cheng and Kang, Iu-thing and Li, Shang-Wen and Lee, Hung-yi},
  journal={IEEE/ACM Transactions on Audio, Speech, and Language Processing},
  year={2024},
  publisher={IEEE}
}

@inproceedings{kuan2026,
  title={AQUA-Bench: Beyond Finding Answers to Knowing When There Are None in Audio Question Answering},
  author={Kuan, Chun-Yi and Lee, Hung-yi},
  booktitle={ICASSP 2026-2026 IEEE International Conference on Acoustics, Speech and Signal Processing (ICASSP)},
  pages={1--5},
  year={2026},
}

@article{kuan2026aqascore,
  title={AQAScore: Evaluating Semantic Alignment in Text-to-Audio Generation via Audio Question Answering},
  author={Kuan, Chun-Yi and Chang, Kai-Wei and Lee, Hung-yi},
  journal={arXiv preprint arXiv:2601.14728},
  year={2026}
}

@inproceedings{alhamoud2025vision,
  title={Vision-language models do not understand negation},
  author={Alhamoud, Kumail and Alshammari, Shaden and Tian, Yonglong and Li, Guohao and Torr, Philip HS and Kim, Yoon and Ghassemi, Marzyeh},
  booktitle={Proceedings of the Computer Vision and Pattern Recognition Conference},
  pages={29612--29622},
  year={2025}
}

@phdthesis{laka1990negation,
  title={Negation in syntax--on the nature of functional categories and projections},
  author={Laka Mugarza, Miren Itziar},
  year={1990},
  school={Massachusetts Institute of Technology}
}

@article{zhang2025negvqa,
  title={NegVQA: Can Vision Language Models Understand Negation?},
  author={Zhang, Yuhui and Su, Yuchang and Liu, Yiming and Yeung-Levy, Serena},
  journal={arXiv preprint arXiv:2505.22946},
  year={2025}
}

@inproceedings{seeing_negation_2026,
  title={Seeing What’s Not There: Negation Understanding Needs More Than Training},
  author={Aggarwal, Bhuvan and More, Amit and Soni, Mudit and Bhat, S Divakar},
  booktitle={The Fourteenth International Conference on Learning Representations},
  year={2026}
}

@inproceedings{sakshi2024mmau,
  title={MMAU: A Massive Multi-Task Audio Understanding and Reasoning Benchmark},
  author={Sakshi, S and Tyagi, Utkarsh and Kumar, Sonal and Seth, Ashish and Selvakumar, Ramaneswaran and Nieto, Oriol and Duraiswami, Ramani and Ghosh, Sreyan and Manocha, Dinesh},
  booktitle={The Thirteenth International Conference on Learning Representations},
  year={2025}
}

@article{arora2025landscape,
  title={On The Landscape of Spoken Language Models: A Comprehensive Survey},
  author={Arora, Siddhant and Chang, Kai-Wei and Chien, Chung-Ming and Peng, Yifan and Wu, Haibin and Adi, Yossi and Dupoux, Emmanuel and Lee, Hung-yi and Livescu, Karen and Watanabe, Shinji},
  journal={Transactions on Machine Learning Research},
  issn={2835-8856},
  year={2025},
}

@article{pai2025billy,
  title={Billy: Steering large language models via merging persona vectors for creative generation},
  author={Pai, Tsung-Min and Wang, Jui-I and Lu, Li-Chun and Sun, Shao-Hua and Lee, Hung-Yi and Chang, Kai-Wei},
  journal={arXiv preprint arXiv:2510.10157},
  year={2025}
}

@article{xie2025emosteer,
  title={EmoSteer-TTS: Fine-Grained and Training-Free Emotion-Controllable Text-to-Speech via Activation Steering},
  author={Xie, Tianxin and Yang, Shan and Li, Chenxing and Yu, Dong and Liu, Li},
  journal={arXiv preprint arXiv:2508.03543},
  year={2025}
}

@article{chen2025persona,
  title={Persona vectors: Monitoring and controlling character traits in language models},
  author={Chen, Runjin and Arditi, Andy and Sleight, Henry and Evans, Owain and Lindsey, Jack},
  journal={arXiv preprint arXiv:2507.21509},
  year={2025}
}

@article{yang2025paras2s,
  title={ParaS2S: Benchmarking and Aligning Spoken Language Models for Paralinguistic-aware Speech-to-Speech Interaction},
  author={Yang, Shu-wen and Tu, Ming and Liu, Andy T and Qu, Xinghua and Lee, Hung-yi and Lu, Lu and Wang, Yuxuan and Wu, Yonghui},
  journal={arXiv preprint arXiv:2511.08723},
  year={2025}
}

@article{ji2024wavchat,
  title={Wavchat: A survey of spoken dialogue models},
  author={Ji, Shengpeng and Chen, Yifu and Fang, Minghui and Zuo, Jialong and Lu, Jingyu and Wang, Hanting and Jiang, Ziyue and Zhou, Long and Liu, Shujie and Cheng, Xize and others},
  journal={arXiv preprint arXiv:2411.13577},
  year={2024}
}

@article{peng2025survey,
  title={A survey on speech large language models for understanding},
  author={Peng, Jing and Wang, Yucheng and Li, Bohan and Guo, Yiwei and Wang, Hankun and Fang, Yangui and Xi, Yu and Li, Haoyu and Li, Xu and Zhang, Ke and others},
  journal={IEEE Journal of Selected Topics in Signal Processing},
  year={2025},
  publisher={IEEE}
}

@inproceedings{cui2025recent,
  title={Recent advances in speech language models: A survey},
  author={Cui, Wenqian and Yu, Dianzhi and Jiao, Xiaoqi and Meng, Ziqiao and Zhang, Guangyan and Wang, Qichao and Guo, Steven Y and King, Irwin},
  booktitle={Proceedings of the 63rd Annual Meeting of the Association for Computational Linguistics (Volume 1: Long Papers)},
  pages={13943--13970},
  year={2025}
}

@article{mousavidiscrete,
  title={Discrete Audio Tokens: More Than a Survey!},
  author={Mousavi, Pooneh and Maimon, Gallil and Moumen, Adel and Petermann, Darius and Shi, Jiatong and Wu, Haibin and Yang, Haici and Kuznetsova, Anastasia and Ploujnikov, Artem and Marxer, Ricard and others},
  journal={Transactions on Machine Learning Research},
  year={2025}
}

@article{wu2024towards,
  title={Towards audio language modeling--an overview},
  author={Wu, Haibin and Chen, Xuanjun and Lin, Yi-Cheng and Chang, Kai-wei and Chung, Ho-Lam and Liu, Alexander H and Lee, Hung-yi},
  journal={arXiv preprint arXiv:2402.13236},
  year={2024}
}

@article{guo2025recent,
  title={Recent advances in discrete speech tokens: A review},
  author={Guo, Yiwei and Li, Zhihan and Wang, Hankun and Li, Bohan and Shao, Chongtian and Zhang, Hanglei and Du, Chenpeng and Chen, Xie and Liu, Shujie and Yu, Kai},
  journal={IEEE Transactions on Pattern Analysis and Machine Intelligence},
  year={2025},
  publisher={IEEE}
}

@article{zhou2026emoshift,
  title={EmoShift: Lightweight Activation Steering for Enhanced Emotion-Aware Speech Synthesis},
  author={Zhou, Li and Jiang, Hao and Li, Junjie and Wang, Tianrui and Li, Haizhou},
  journal={arXiv preprint arXiv:2601.22873},
  year={2026}
}

@article{facchiano2025activation,
  title={Activation patching for interpretable steering in music generation},
  author={Facchiano, Simone and Strano, Giorgio and Crisostomi, Donato and Tallini, Irene and Mencattini, Tommaso and Galasso, Fabio and Rodol{\`a}, Emanuele},
  journal={arXiv preprint arXiv:2504.04479},
  year={2025}
}

@article{kang2025model,
  title={Model Whisper: Steering Vectors Unlock Large Language Models' Potential in Test-time},
  author={Kang, Xinyue and Shi, Diwei and Chen, Li},
  journal={arXiv preprint arXiv:2512.04748},
  year={2025}
}

@article{tang2025wave,
  title={WAVE: Learning Unified \& Versatile Audio-Visual Embeddings with Multimodal LLM},
  author={Tang, Changli and Xiao, Qinfan and Mei, Ke and Wang, Tianyi and Rao, Fengyun and Zhang, Chao},
  journal={arXiv preprint arXiv:2509.21990},
  year={2025}
}

@article{kuan2026walking,
  title={Walking Through Uncertainty: An Empirical Study of Uncertainty Estimation for Audio-Aware Large Language Models},
  author={Kuan, Chun-Yi and Huang, Wei-Ping and Lee, Hung-yi},
  journal={arXiv preprint arXiv:2604.25591},
  year={2026}
}

@inproceedings{yuksekgonuland,
  title={When and Why Vision-Language Models Behave like Bags-Of-Words, and What to Do About It?},
  author={Yuksekgonul, Mert and Bianchi, Federico and Kalluri, Pratyusha and Jurafsky, Dan and Zou, James},
  booktitle={The Eleventh International Conference on Learning Representations},
  year={2023}
}

@inproceedings{thrush2022winoground,
  title={Winoground: Probing vision and language models for visio-linguistic compositionality},
  author={Thrush, Tristan and Jiang, Ryan and Bartolo, Max and Singh, Amanpreet and Williams, Adina and Kiela, Douwe and Ross, Candace},
  booktitle={Proceedings of the IEEE/CVF Conference on Computer Vision and Pattern Recognition},
  pages={5238--5248},
  year={2022}
}

@inproceedings{ma2023crepe,
  title={Crepe: Can vision-language foundation models reason compositionally?},
  author={Ma, Zixian and Hong, Jerry and Gul, Mustafa Omer and Gandhi, Mona and Gao, Irena and Krishna, Ranjay},
  booktitle={Proceedings of the IEEE/CVF Conference on Computer Vision and Pattern Recognition},
  pages={10910--10921},
  year={2023}
}

@article{hsieh2023sugarcrepe,
  title={Sugarcrepe: Fixing hackable benchmarks for vision-language compositionality},
  author={Hsieh, Cheng-Yu and Zhang, Jieyu and Ma, Zixian and Kembhavi, Aniruddha and Krishna, Ranjay},
  journal={Advances in neural information processing systems},
  volume={36},
  pages={31096--31116},
  year={2023}
}

@article{zhao2022vl,
  title={Vl-checklist: Evaluating pre-trained vision-language models with objects, attributes and relations},
  author={Zhao, Tiancheng and Zhang, Tianqi and Zhu, Mingwei and Shen, Haozhan and Lee, Kyusong and Lu, Xiaopeng and Yin, Jianwei},
  journal={arXiv preprint arXiv:2207.00221},
  year={2022}
}

@inproceedings{lewis2024does,
  title={Does clip bind concepts? probing compositionality in large image models},
  author={Lewis, Martha and Nayak, Nihal and Yu, Peilin and Merullo, Jack and Yu, Qinan and Bach, Stephen and Pavlick, Ellie},
  booktitle={Findings of the Association for Computational Linguistics: EACL 2024},
  pages={1487--1500},
  year={2024}
}

@inproceedings{sammani2026negation,
  title={When Negation Is a Geometry Problem in Vision Language Models},
  author={Sammani, Fawaz and Chamiti, Tzoulio and Gavrikov, Paul and Deligiannis, Nikos},
  booktitle={Proceedings of the IEEE/CVF Conference on Computer Vision and Pattern Recognition},
  pages={11553--11562},
  year={2026}
}

@inproceedings{xu2026omni,
  title={Omni-NegCLIP: Enhancing CLIP with Front-Layer Contrastive Fine-Tuning for Comprehensive Negation Understanding},
  author={Xu, Jingqi},
  booktitle={Proceedings of the IEEE/CVF Conference on Computer Vision and Pattern Recognition},
  pages={9392--9401},
  year={2026}
}

@article{chang2026tico,
  title={TiCo: Time-Controllable Training for Spoken Dialogue Models},
  author={Chang, Kai-Wei and Chen, Wei-Chih and Hu, En-Pei and Lee, Hung-yi and Glass, James},
  journal={arXiv preprint arXiv:2603.22267},
  year={2026}
}

@inproceedings{takano2025human,
  title={Human-clap: Human-perception-based contrastive language-audio pretraining},
  author={Takano, Taisei and Okamoto, Yuki and Kanamori, Yusuke and Saito, Yuki and Nagase, Ryotaro and Saruwatari, Hiroshi},
  booktitle={2025 Asia Pacific Signal and Information Processing Association Annual Summit and Conference (APSIPA ASC)},
  pages={131--136},
  year={2025},
  organization={IEEE}
}

@inproceedings{dinkel2026glap,
  title={GLAP: General contrastive audio-text pretraining across domains and languages},
  author={Dinkel, Heinrich and Yan, Zhiyong and Wang, Tianzi and Wang, Yongqing and Sun, Xingwei and Niu, Yadong and Liu, Jizhong and Li, Gang and Zhang, Junbo and Luan, Jian},
  booktitle={ICASSP 2026-2026 IEEE International Conference on Acoustics, Speech and Signal Processing (ICASSP)},
  pages={14737--14741},
  year={2026},
  organization={IEEE}
}

@article{tseng2025revisiting,
  title={Revisiting Audio-language Pretraining for Learning General-purpose Audio Representation},
  author={Tseng, Wei-Cheng and Zhou, Xuanru and Huo, Mingyue and Shao, Yiwen and Zhang, Hao and Yu, Dong},
  journal={arXiv preprint arXiv:2511.16757},
  year={2025}
}

@article{jing2024paraclap,
  title={ParaCLAP--Towards a general language-audio model for computational paralinguistic tasks},
  author={Jing, Xin and Triantafyllopoulos, Andreas and Schuller, Bj{\"o}rn},
  journal={arXiv preprint arXiv:2406.07203},
  year={2024}
}

@inproceedings{ye2023clapspeech,
  title={Clapspeech: Learning prosody from text context with contrastive language-audio pre-training},
  author={Ye, Zhenhui and Huang, Rongjie and Ren, Yi and Jiang, Ziyue and Liu, Jinglin and He, Jinzheng and Yin, Xiang and Zhao, Zhou},
  booktitle={Proceedings of the 61st Annual Meeting of the Association for Computational Linguistics (Volume 1: Long Papers)},
  pages={9317--9331},
  year={2023}
}




\clearpage
\appendices
\section{Controlled Diagnosis of Text-Side Negation Handling}
\label{app:controlled}

This appendix provides the implementation details and full results for the
controlled diagnosis in Section~\ref{sec:textdiag}. The main paper reports the
results averaged over five LAION-CLAP~\cite{wu2023large} checkpoints; here, we additionally include
the LLM-based embedding model WAVE-7B~\cite{tang2025wave}.

\subsection{Controlled setup}

We construct 1,000 controlled target--distractor pairs from ESC-50
\cite{piczak2015dataset}. 
Each item contains two positive sound events,
$p_1$ and $p_2$, and one designated negated concept $Y$. The two positive
events are concatenated into a 10\,s base clip. 
We then overlay either $Y$ or a
filler event $F$ at the same onset and gain:
\begin{itemize}
  \item \textbf{target}: the base clip with filler $F$, but without $Y$;
  \item \textbf{distractor}: the same base clip with $Y$.
\end{itemize}
The filler $F$ is matched to $Y$ in coarse ESC-50 category and loudness.
Therefore, the target and distractor differ only in whether the overlaid event
is $F$ or the negated concept $Y$. 
Sound presence is known by construction and does not require ALLM verification.
For each item, we construct a negated query, ``$P$, with no $Y$'', its
affirmative counterpart, ``$P$, with $Y$'', and the oracle description ``$P$'', where $P$ contains the two positive concepts.
We additionally construct the following deterministic text references:
\begin{itemize}
  \item a meaning-preserving \textbf{paraphrase}, obtained by reversing the
  positive-concept order and replacing ``no'' with ``without'';
  \item a \textbf{content-swap} reference, obtained by replacing one positive
  concept;
  \item an \textbf{unrelated} query from another item;
  \item three \textbf{scope variants}, which negate each of
  $\{p_1,p_2,Y\}$ in turn;
  \item a \textbf{bag-of-concepts} reference containing only the three concept
  phrases.
\end{itemize}

For WAVE-7B, an instruction-aware MLLM embedding model, audio embeddings are conditioned on a text instruction. 
We use ``Please describe the audio.'' as the instruction.
All embeddings are $L_2$-normalized before computing
cosine similarity.

\subsection{Metrics}

Let $q$ be the negated-query embedding, and let
$t_{\mathrm{aff}}$, $t_{\mathrm{para}}$, $t_{\mathrm{swap}}$, and
$t_{\mathrm{unrel}}$ denote the affirmative, paraphrase, content-swap, and
unrelated references, respectively.

\medskip
\noindent\textbf{Polarity sensitivity.}
We report
\begin{align}
\text{normalized affinity} &=
\frac{s(q,t_{\mathrm{aff}})-s(q,t_{\mathrm{unrel}})}
     {s(q,t_{\mathrm{para}})-s(q,t_{\mathrm{unrel}})},\\
\text{polarity}/\text{content} &=
\frac{1-s(q,t_{\mathrm{aff}})}
     {1-s(q,t_{\mathrm{swap}})}.
\end{align}
A normalized affinity near or above $1$ means that the truth-opposite
affirmative is as close to the query as its meaning-preserving paraphrase. A
polarity/content ratio below $1$ means that changing polarity moves the
embedding less than replacing one sound concept.

\medskip
\noindent\textbf{Negation scope.}
For the three scope variants $\{v_1,v_2,v_3\}$, we compute
\[
\text{scope-perm}
=
\mathrm{mean}_{i<j}\,s(v_i,v_j).
\]
We compare this value with the content-swap similarity
$s(q,t_{\mathrm{swap}})$ and the bag-of-concepts similarity
$s(q,t_{\mathrm{bag}})$. High scope-permutation and bag-of-concepts
similarities indicate weak encoding of which concept is negated.

\medskip
\noindent\textbf{Audio-retrieval bridge.}
For each item, we define
\[
m(q)=s(q,a_{\mathrm{dist}})-s(q,a_{\mathrm{tgt}}).
\]
A positive margin indicates that the query prefers the distractor containing
the negated concept. We report the mean margins for the negated and affirmative
queries, together with their item-level correlation.


\subsection{Full results}

\begin{table}[t]
\centering
\caption{
Controlled text-side negation diagnosis for the mean over five LAION-CLAP
checkpoints and WAVE-7B.
}
\label{tab:controlled_diagnosis_sup}

\small
\setlength{\tabcolsep}{3pt}
\renewcommand{\arraystretch}{1.05}

\begin{tabularx}{\columnwidth}{
  @{}
  >{\raggedright\arraybackslash}X
  c
  c
  @{}
}
\toprule
\textbf{Diagnostic} & \textbf{CLAP} & \textbf{WAVE-7B} \\
\midrule

\multicolumn{3}{@{}l}{\textit{Polarity sensitivity}} \\
Normalized affinity to affirmative text & 1.07 & 1.05 \\
Polarity change / content swap          & 0.30 & 0.12 \\

\midrule
\multicolumn{3}{@{}l}{\textit{Negation scope}} \\
Scope-permutation similarity & 0.70 & 0.90 \\
Content-swap similarity      & 0.58 & 0.70 \\
Bag-of-concepts similarity   & 0.82 & 0.91 \\

\midrule
\multicolumn{3}{@{}l}{\textit{Audio-retrieval bridge}} \\
Negated--affirmative margin correlation
    & 0.76 & 0.94 \\
Negated-query margin
    & +0.018 & +0.016 \\
Affirmative-query margin
    & +0.072 & +0.022 \\

\bottomrule
\end{tabularx}
\end{table}

Table~\ref{tab:controlled_diagnosis_sup} reveals three consistent patterns
across both model families.

\noindent\textbf{Negation is only a weak perturbation.}
The affirmative counterpart is at least as close to the negated query as the
meaning-preserving paraphrase, with normalized affinities of $1.07$ for CLAP
and $1.05$ for WAVE-7B. 
Polarity changes are also substantially smaller than
content swaps, particularly for WAVE-7B, whose polarity/content ratio is only
$0.12$. Thus, changing whether a concept is affirmed or negated affects the
embedding much less than changing the concept itself.

\noindent\textbf{Negation is weakly bound to the intended concept.}
For CLAP, the scope-permutation similarity ($0.70$) exceeds the content-swap
similarity ($0.58$), while the bag-of-concepts similarity remains high
($0.82$). The pattern is even stronger for WAVE-7B: its scope variants have a
similarity of $0.90$, close to its bag-of-concepts similarity of $0.91$.
These results indicate that the representations preserve which sound concepts
are mentioned more strongly than which concept is negated.

\noindent\textbf{The text-side failure carries over to audio ranking.}
Negated and affirmative queries induce strongly correlated audio margins
($0.76$ for CLAP and $0.94$ for WAVE-7B). Moreover, the negated-query margins
remain positive ($+0.018$ and $+0.016$, respectively), indicating continued
preference for the audio containing the forbidden concept. Negated queries
therefore behave like attenuated affirmative queries rather than reversing the
ranking toward the target audio.

\noindent\textbf{Overall takeaway.}
Across both model families, negation is represented as a weak modifier of
mentioned concepts rather than an exclusion operator. This behavior persists
from text representations to cross-modal audio ranking and is, on several
measures, even more pronounced for WAVE-7B.

\section{Full $\lambda$-Sweep for Training-Free Steering}
\label{steering-lamda-sweep}
\begin{table*}[hbpt]
\centering
\caption{
Retrieval-Neg (R@5) versus steering strength $\lambda$ on \textbf{AudioCaps test}~\cite{kim2019audiocaps}.
Per-model best results are shown in bold, the value used in the main paper
($\lambda=0.2$) is underlined, and gray cells denote the no-steering baseline
($\lambda=0$).
}
\small
\begin{tabular}{l cccccccccccc}
\toprule
Model & \cellcolor{lightgray}0 & 0.1 & 0.2 & 0.3 & 0.5 & 0.7 & 1 & 1.5 & 2 & 3 & 5 & 8 \\
\midrule
\textit{LAION-CLAP} \\
-- 630k-audioset-best 
& \cellcolor{lightgray}60.3 & 61.5 & \underline{\textbf{61.8}} & 60.8 & 55.4 & 48.6 & 36.8 & 20.9 & 10.8 & 4.2 & 1.9 & 0.9 \\
-- 630k-best 
& \cellcolor{lightgray}59.1 & \textbf{59.5} & \underline{58.5} & 57.3 & 51.1 & 43.0 & 27.0 & 12.8 & 6.5 & 2.4 & 0.4 & 0.1 \\
-- music-audioset 
& \cellcolor{lightgray}57.0 & 57.2 & \underline{\textbf{57.3}} & 55.9 & 50.8 & 44.8 & 33.4 & 18.7 & 10.4 & 3.8 & 2.0 & 0.8 \\
-- music-speech-audioset 
& \cellcolor{lightgray}57.9 & 58.4 & \underline{58.6} & \textbf{59.0} & 54.4 & 49.1 & 39.0 & 21.7 & 13.5 & 6.0 & 2.8 & 1.2 \\
-- music-speech 
& \cellcolor{lightgray}59.4 & \textbf{59.8} & \underline{58.1} & 56.8 & 53.1 & 44.9 & 28.8 & 12.8 & 6.2 & 2.4 & 0.6 & 0.1 \\
\textit{M2D-CLAP} \\
-- M2D-CLAP-2025 
& \cellcolor{lightgray}63.2 & 64.9 & \underline{65.8} & 66.8 & \textbf{67.1} & 64.7 & 58.4 & 44.6 & 30.8 & 13.7 & 4.5 & 1.6 \\
-- M2D-CLAP-2024 
& \cellcolor{lightgray}46.1 & 49.0 & \underline{50.2} & \textbf{50.9} & 48.8 & 41.9 & 29.7 & 14.8 & 7.7 & 3.6 & 1.6 & 0.9 \\
\textit{WAVE} \\
-- WAVE-7B 
& \cellcolor{lightgray}75.2 & 75.8 & \underline{77.1} & \textbf{77.2} & 76.1 & 73.4 & 68.4 & 53.0 & 39.0 & 19.4 & 4.8 & 2.3 \\
\bottomrule
\end{tabular}
\label{tab:supp_retrieval_audiocaps}
\end{table*}

\begin{table*}[htbp]
\centering
\caption{
Retrieval-Neg (R@5) versus steering strength $\lambda$ on \textbf{Clotho evaluation}~\cite{drossos2020clotho}.
Per-model best results are shown in bold, the value used in the main paper
($\lambda=0.2$) is underlined, and gray cells denote the no-steering baseline
($\lambda=0$).
}
\small
\begin{tabular}{l cccccccccccc}
\toprule
Model & \cellcolor{lightgray}0 & 0.1 & 0.2 & 0.3 & 0.5 & 0.7 & 1 & 1.5 & 2 & 3 & 5 & 8 \\
\midrule
\textit{LAION-CLAP} \\
-- 630k-audioset-best 
& \cellcolor{lightgray}35.4 & 35.9 & \underline{\textbf{36.7}} & 36.3 & 34.8 & 30.0 & 22.2 & 11.4 & 7.1 & 2.7 & 1.2 & 0.9 \\
-- 630k-best 
& \cellcolor{lightgray}37.0 & 37.9 & \underline{\textbf{38.7}} & 38.2 & 36.2 & 31.8 & 24.4 & 13.8 & 9.3 & 3.6 & 1.5 & 0.9 \\
-- music-audioset 
& \cellcolor{lightgray}33.8 & 34.2 & \underline{\textbf{35.0}} & 34.6 & 30.2 & 27.6 & 21.9 & 13.0 & 7.5 & 3.6 & 1.8 & 1.3 \\
-- music-speech-audioset 
& \cellcolor{lightgray}34.2 & \textbf{35.5} & \underline{34.8} & 34.6 & 32.3 & 27.7 & 20.8 & 11.1 & 6.0 & 3.1 & 1.5 & 0.7 \\
-- music-speech 
& \cellcolor{lightgray}38.0 & 36.6 & \underline{\textbf{37.5}} & 36.2 & 34.5 & 32.0 & 26.7 & 16.0 & 8.2 & 3.6 & 1.8 & 1.1 \\
\textit{M2D-CLAP} \\
-- M2D-CLAP-2025 
& \cellcolor{lightgray}38.0 & 40.4 & \underline{42.3} & 42.9 & \textbf{43.5} & 41.8 & 35.4 & 25.3 & 17.4 & 8.4 & 3.1 & 1.5 \\
-- M2D-CLAP-2024 
& \cellcolor{lightgray}19.6 & 20.9 & \underline{22.5} & \textbf{22.9} & 22.8 & 20.6 & 14.4 & 8.8 & 5.4 & 2.8 & 0.8 & 0.5 \\
\textit{WAVE} \\
-- WAVE-7B 
& \cellcolor{lightgray}50.0 & 51.6 & \underline{\textbf{52.0}} & 51.9 & 51.1 & 49.8 & 46.5 & 38.3 & 28.8 & 14.4 & 5.4 & 2.4 \\
\bottomrule
\end{tabular}
\label{tab:supp_retrieval_clotho}
\end{table*}

\begin{table*}[htbp]
\centering
\caption{
MCQ-Neg (Avg) vs.\ steering strength $\lambda$ on \textbf{AudioCaps test}~\cite{kim2019audiocaps}.
Per-model \textbf{best} results are bold, $\lambda{=}0.2$ is
\underline{underlined}, and the no-steering baseline ($\lambda{=}0$) is
shaded gray.
}
\small
\begin{tabular}{l cccccccccccc}
\toprule
Model & \cellcolor{lightgray}0 & 0.1 & 0.2 & 0.3 & 0.5 & 0.7 & 1 & 1.5 & 2 & 3 & 5 & 8 \\
\midrule
\textit{LAION-CLAP} \\
-- 630k-audioset-best 
& \cellcolor{lightgray}43.2 & 44.6 & \underline{45.0} & \textbf{45.4} & 44.9 & 42.6 & 39.1 & 36.8 & 36.1 & 35.0 & 34.7 & 34.5 \\
-- 630k-best 
& \cellcolor{lightgray}41.9 & 42.7 & \underline{\textbf{42.8}} & 42.1 & 39.9 & 38.3 & 36.3 & 34.7 & 34.4 & 33.9 & 33.9 & 33.8 \\
-- music-audioset 
& \cellcolor{lightgray}43.3 & \textbf{44.9} & \underline{44.9} & 44.7 & 42.5 & 40.6 & 39.3 & 37.5 & 36.3 & 35.0 & 35.0 & 34.6 \\
-- music-speech-audioset 
& \cellcolor{lightgray}41.1 & 42.0 & \underline{\textbf{42.2}} & 41.1 & 41.0 & 41.2 & 40.8 & 38.4 & 36.9 & 36.0 & 35.2 & 34.9 \\
-- music-speech 
& \cellcolor{lightgray}34.9 & 36.5 & \underline{38.4} & \textbf{40.0} & 39.7 & 38.6 & 36.8 & 34.9 & 34.4 & 34.0 & 33.9 & 33.7 \\
\textit{M2D-CLAP} \\
-- M2D-CLAP-2025 
& \cellcolor{lightgray}27.9 & 29.2 & \underline{30.2} & 31.6 & 34.4 & 35.9 & \textbf{36.2} & 34.7 & 33.9 & 33.6 & 33.4 & 33.4 \\
-- M2D-CLAP-2024 
& \cellcolor{lightgray}30.8 & 31.3 & \underline{30.9} & 31.0 & 31.7 & 31.9 & 31.7 & 31.4 & 31.7 & 31.9 & 32.6 & \textbf{33.0} \\
\textit{WAVE} \\
-- WAVE-7B 
& \cellcolor{lightgray}43.5 & 51.8 & \underline{\textbf{54.5}} & 50.8 & 40.1 & 34.7 & 33.6 & 33.3 & 33.3 & 33.3 & 33.3 & 33.3 \\
\bottomrule
\end{tabular}
\label{tab:supp_mcq_audiocaps}
\end{table*}

\begin{table*}[htbp]
\centering
\caption{
MCQ-Neg (Avg) vs.\ steering strength $\lambda$ on \textbf{Clotho evaluation}~\cite{drossos2020clotho}.
Per-model \textbf{best} results are bold, $\lambda{=}0.2$ is
\underline{underlined}, and the no-steering baseline ($\lambda{=}0$) is
shaded gray.
}
\small
\begin{tabular}{l cccccccccccc}
\toprule
Model & \cellcolor{lightgray}0 & 0.1 & 0.2 & 0.3 & 0.5 & 0.7 & 1 & 1.5 & 2 & 3 & 5 & 8 \\
\midrule
\textit{LAION-CLAP} \\
-- 630k-audioset-best 
& \cellcolor{lightgray}37.2 & 37.0 & \underline{36.6} & 36.8 & \textbf{37.8} & 37.8 & 36.6 & 35.4 & 34.8 & 34.3 & 34.4 & 34.2 \\
-- 630k-best 
& \cellcolor{lightgray}38.1 & 39.2 & \underline{39.4} & \textbf{39.6} & 38.8 & 38.1 & 36.3 & 35.0 & 34.4 & 34.4 & 34.0 & 33.9 \\
-- music-audioset 
& \cellcolor{lightgray}38.4 & \textbf{38.9} & \underline{37.7} & 37.1 & 36.8 & 36.7 & 35.5 & 34.6 & 34.4 & 33.9 & 33.7 & 33.7 \\
-- music-speech-audioset 
& \cellcolor{lightgray}30.9 & 31.0 & \underline{31.1} & 31.3 & 32.4 & 33.0 & 32.9 & 32.8 & 32.8 & 33.2 & \textbf{33.6} & 33.5 \\
-- music-speech 
& \cellcolor{lightgray}27.9 & 29.4 & \underline{29.5} & 30.0 & 32.1 & 33.1 & 34.0 & \textbf{34.5} & 34.0 & 34.4 & 34.1 & 33.7 \\
\textit{M2D-CLAP} \\
-- M2D-CLAP-2025 
& \cellcolor{lightgray}22.1 & 22.2 & \underline{22.9} & 24.1 & 26.2 & 27.2 & 29.4 & 31.9 & 32.3 & 33.2 & 33.4 & \textbf{33.5} \\
-- M2D-CLAP-2024 
& \cellcolor{lightgray}23.2 & 23.6 & \underline{24.1} & 24.3 & 25.5 & 26.8 & 28.0 & 29.6 & 30.4 & 31.1 & 31.6 & \textbf{31.9} \\
\textit{WAVE} \\
-- WAVE-7B 
& \cellcolor{lightgray}42.2 & 49.3 & \underline{\textbf{52.4}} & 51.2 & 43.5 & 36.9 & 34.1 & 33.4 & 33.3 & 33.3 & 33.3 & 33.3 \\
\bottomrule
\end{tabular}
\label{tab:supp_mcq_clotho}
\end{table*}

\begin{figure*}[ht]
    \centering
    \includegraphics[width=1.0\textwidth]{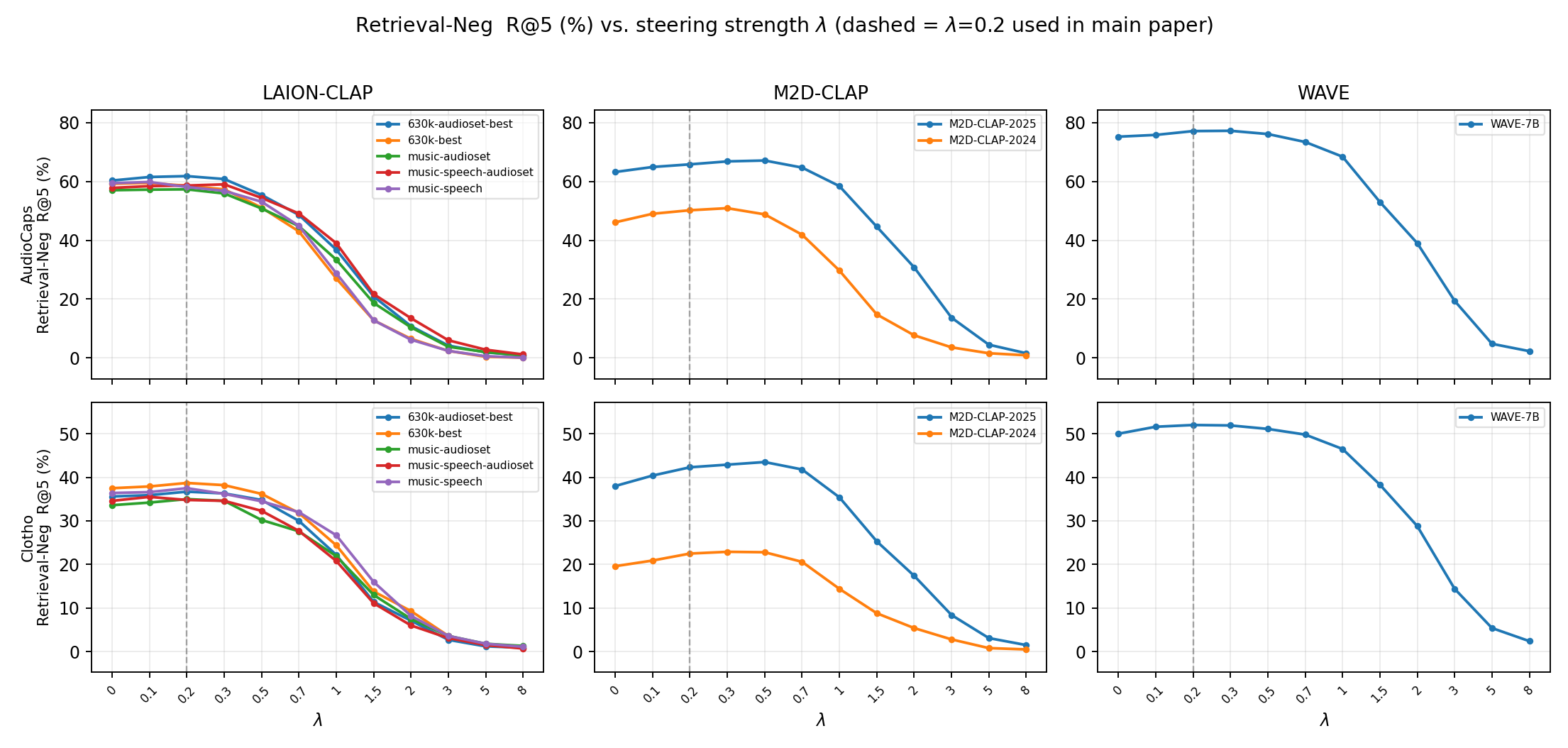} 
    \caption{
Effect of steering strength $\lambda$ on Retrieval-Neg R@5 for AudioCaps~\cite{kim2019audiocaps}
(top) and Clotho~\cite{drossos2020clotho} (bottom). 
Performance generally peaks at a small or
moderate $\lambda$ and declines sharply under stronger steering, consistent with excessive subtraction removing useful query semantics.
}
    \label{fig:steering-retrieval}
\end{figure*}

\begin{figure*}[ht]
    \centering
    \includegraphics[width=1.0\textwidth]{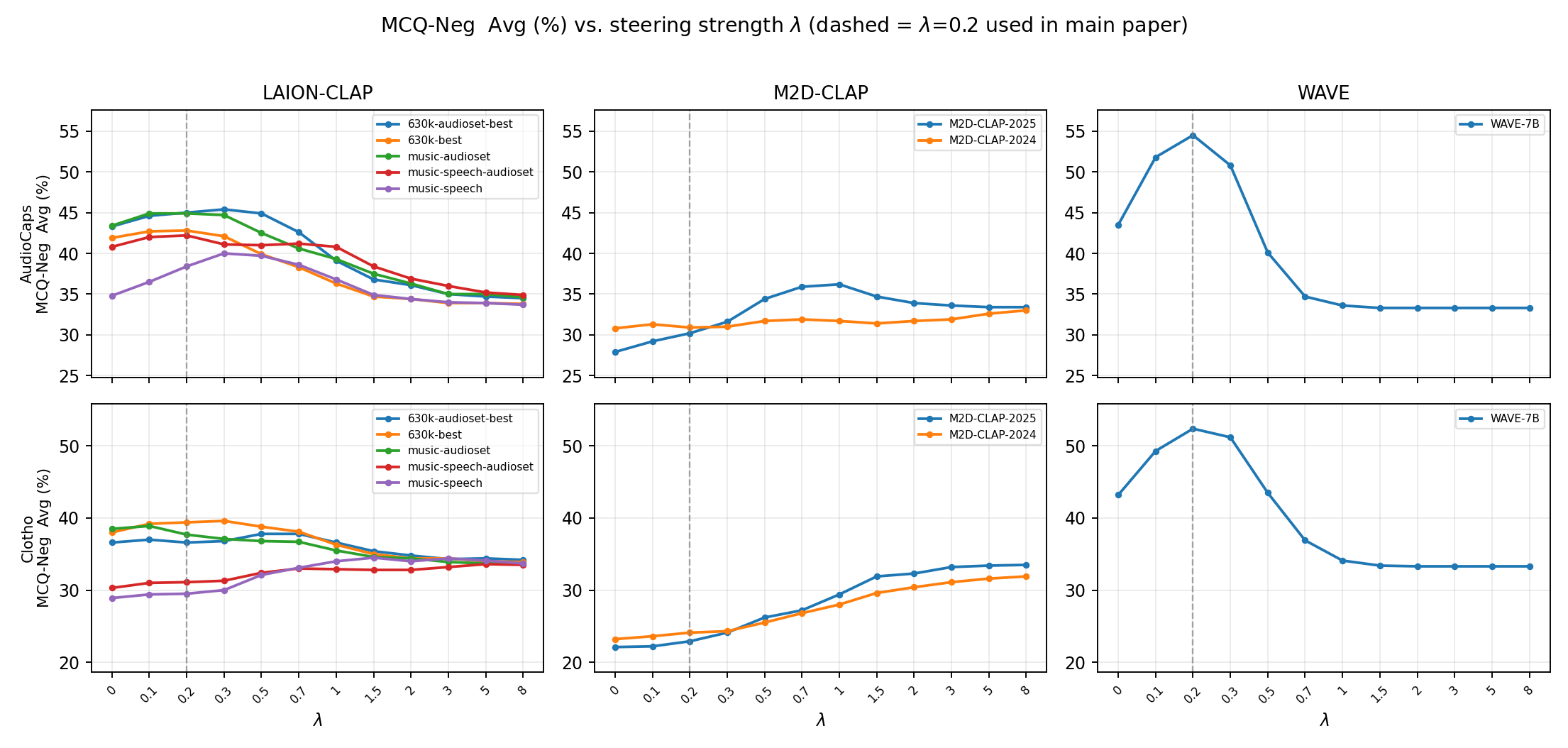} 
    \caption{
MCQ-Neg performance across steering strengths $\lambda$ on AudioCaps~\cite{kim2019audiocaps} (top) and Clotho~\cite{drossos2020clotho} (bottom). 
Steering improves local option ranking for several
models, but the optimal strength varies across models and datasets. 
This suggests that inference-time steering can partially modify negation-related decisions, although the gains are not consistently robust.
}
    \label{fig:steering-mcq}
\end{figure*}

To complement the main results, we report the full $\lambda$-sweep of training-free negation steering in~\Cref{tab:supp_retrieval_audiocaps,tab:supp_retrieval_clotho,tab:supp_mcq_audiocaps,tab:supp_mcq_clotho}, with corresponding trends visualized in~\Cref{fig:steering-retrieval,fig:steering-mcq}.
The sweep shows that small steering strengths can improve negation-aware matching, but the gains are limited and model-dependent.
In Retrieval-Neg, performance usually peaks at a small or moderate $\lambda$ and then drops sharply as $\lambda$ increases, indicating that excessive steering removes not only the negated concept but also useful semantic information from the query. 
In MCQ-Neg, some models benefit more from larger $\lambda$ values, especially when the task only requires local ranking among a small set of options. 
These trends support our main conclusion: training-free steering can partially edit local decisions, but it does not provide a robust solution to negation failure in the global retrieval space.

\section{On the Use of ALLMs for Verification}

A potential concern is circularity, as we use an audio-aware LLM (ALLM) to construct a benchmark for diagnosing negation failures in audio-language models. We argue that this design does not compromise the validity of NegEval-Audio for three reasons.
First, the verification task is purely affirmative. For each candidate concept, the ALLM is asked, ``Does the audio contain the sound of X?'' 
This is a presence-detection task that does not require interpreting a negated query. 
Negation is introduced only after verification, when the resulting concept sets are used to construct retrieval queries and MCQ options. 
Thus, the capability being tested by NegEval-Audio is not required during benchmark construction.

Second, the verifier and the evaluated models differ in both architecture and inference paradigm. 
The evaluated models are embedding-based systems that rank audio--text pairs according to similarity in a shared representation space, whereas the verifier is a generative ALLM that produces a binary answer. 
Although this distinction alone does not guarantee independent failure modes, it reduces the concern that the benchmark construction procedure directly inherits the specific negation-related ranking failures being evaluated.

Most importantly, the verification results are independently validated by human annotators. 
Across 200 audio clips and 460 verified negative concepts, human judgments agree with the ALLM verification results in 96.7\% of cases. 
This high agreement provides direct evidence that the resulting negative concepts are reliable.

\end{document}